\documentclass[preprint,12pt]{elsarticle}

\makeatletter
\def\ps@pprintTitle{%
  \let\@oddhead\@empty
  \let\@evenhead\@empty
  \let\@oddfoot\@empty
  \let\@evenfoot\@empty
}
\makeatother
\usepackage{amssymb}
\usepackage{amsmath}
\usepackage{float}
\usepackage{lipsum}
\usepackage{caption}    
\usepackage{subcaption} 
\usepackage{tabularx}
\newcolumntype{C}{>{\centering\arraybackslash}X}
\usepackage{placeins}
\usepackage{enumitem}

\newlist{num}{enumerate}{1}
\setlist[num]{label=\arabic*.}

\begin{document}

\begin{frontmatter}

\title{Development and Characterization of a Time Projection Chamber Prototype for Neutron Oscillation Searches at the European Spallation Source} 

\author[lu]{B. Rataj\corref{cor1}}
\ead{ blahoslav.rataj@fysik.lu.se}
\author[lu,ess]{V. Santoro}
\author[lu]{A. Oskarsson}
\author[lu]{M. Holl}
\author[lu]{V. Hehl}
\author[lu]{D. Silvermyr}

\affiliation[lu]{
  organization={Department of Physics, Lund University},
  addressline={P.O. Box 118},
  postcode={SE-221 00},
  city={Lund},
  country={Sweden}
}

\affiliation[ess]{
  organization={European Spallation Source ERIC},
  addressline={Partikelgatan 5},
  postcode={22484},
  city={Lund},
  country={Sweden}
}

\cortext[cor1]{Corresponding author}

\begin{abstract}
A compact time projection chamber (TPC) prototype was studied for charged-pion tracking in the annihilation detector proposed for the HIBEAM experiment at the European Spallation Source. HIBEAM aims to search for neutron-antineutron oscillations, where a produced antineutron would annihilate on a carbon target and produce a pionic final state. A tracking algorithm was developed and validated using cosmic-particle measurements expected to be dominated by atmospheric cosmic muons, proton elastic-scattering data from the Cyclotron Centre Bronowice, and simulated TPC event data.

Cosmic measurements provided a minimum-ionising reference for pion-like tracks, while proton-scattering data tested reconstruction in a more highly ionising environment. Simulated crossing-track events were used as a stress test of complex topologies. The algorithm performed reliably for well-separated tracks, while crossing or overlapping tracks and dense proton-scattering events reduced performance. In experimental data, reconstructed tracks showed sub-millimetre residual widths, robust charge sharing supporting the zigzag-shaped readout, and uniform measured $dE/dx$ across the drift volume.

The truncated-mean track-level $dE/dx$ distribution in cosmic data showed an approximately Gaussian peak with a relative width of about 16 \%, demonstrating how energy-loss information can be used for particle identification. Tilted proton tracks showed broader residuals in yz, because the pad row signal was extended in time. A sub-centroid refit technique was proposed, using the internal ADC time profile within a pad row instead of a single charge-weighted time position. With four equal-y sub-centroids per pad row, the yz residual width was reduced from 1.30 mm to 0.55 mm. These results validate the initial tracking capabilities of the algorithm and prototype, and provide feedback for the next-generation prototype.
\end{abstract}


\end{frontmatter}

\section{Introduction}

The HIBEAM experiment is proposed as the first stage of the \\ HIBEAM/NNBAR programme at the European Spallation Source (ESS)~\cite{santoro2025hibeam,addazi2021new,Santoro:2024lvc,garoby2017european}. It aims to conduct searches for baryon-number-violating neutron--antineutron (\(n \rightarrow \bar{n}\)) oscillations  using the high-intensity neutron beams available at ESS. A later stage, NNBAR, foresees a larger-scale search. \(n \rightarrow \bar{n}\) oscillation is a hypothetical process that would violate baryon-number conservation by two units~\cite{addazi2021new}. Baryon-number violation is one of the Sakharov conditions required for baryogenesis~\cite{Sakharov:1967dj}, a mechanism that could explain the observed matter--antimatter asymmetry of the Universe. The most recent free-neutron search for \(n \rightarrow \bar{n}\) oscillations was performed at the Institut Laue-Langevin (ILL) in the 1990s and still sets the most stringent direct limit on the free \(n \rightarrow \bar{n}\) oscillation time~\cite{baldo1994new}. Compared with the ILL experiment, HIBEAM is expected to benefit from an improved neutron oscillation sensitivity, higher-resolution calorimetry and three-dimensional charged-particle tracking provided by the TPC~\cite{santoro2025hibeam}.

Free-neutron searches are particularly attractive because they can be performed in a quasi-free regime, where the neutron and antineutron remain nearly degenerate in energy. In this regime, the transition probability scales approximately as
\[
P_{n \rightarrow \bar{n}} \approx \left(\frac{t}{\tau_{n-\bar{n}}}\right)^2,
\]
where \(t\) is the propagation time and \(\tau_{n-\bar{n}}\) is the oscillation time. This makes free-neutron experiments a comparatively clean probe of \(\Delta \mathcal{B}=2\) physics, in contrast to bound-neutron searches where nuclear effects suppress the oscillation probability and introduce model dependencies~\cite{Dover:1982wv,Barrow:2025rhm}.

Given the expected rarity of the process, the experiment must rely on a clean and background-suppressed signature. If an oscillation occurs, the antineutron is expected to annihilate in a thin carbon foil, producing a multi-pion final state with a total energy of approximately \(1.88~\mathrm{GeV}\). This final state contains charged particles and photons from neutral-pion decays. The reconstruction of charged-particle tracks emerging from a common vertex in the annihilation target is therefore one of the central experimental requirements.

In order to identify such events, the HIBEAM annihilation-detector concept combines charged-particle tracking, electromagnetic calorimetry and cosmic-background rejection. The calorimeter is required to detect and measure photons originating from neutral-pion (\(\pi^{0}\)) decays, while the cosmic-ray veto system provides information for the off-line rejection of background events induced by cosmic particles. The focus of this work is the charged particle tracking with the Time Projection Chamber (TPC)~\cite{marx1978time}.

The purpose of this paper is to present the development, characterisation and reconstruction performance of a TPC prototype intended to address the tracking role of the HIBEAM annihilation-detector concept. A TPC is a gaseous detector sensitive to ionising radiation. TPC technology has been widely used in major particle physics experiments, such as the ALICE experiment at CERN~\cite{alme2010alice}, where a large TPC is employed to reconstruct thousands of particle tracks produced in heavy-ion collisions. TPCs have also been successfully adopted in other experiments requiring tracking and particle identification. Although the technology is well established and experimentally validated, the TPC foreseen for HIBEAM must operate under an unique set of constraints.

The detector must be compact and operate without a magnetic field~\cite{santoro2025hibeam}. A magnetic field cannot be applied within the annihilation detector volume, as the neutron beamline is magnetically shielded to preserve the degeneracy between neutron and antineutron states. The presence of a magnetic field would induce an energy splitting between the neutron and antineutron due to their opposite magnetic moments, thereby suppressing the oscillation probability. Consequently, curvature-based momentum reconstruction is not available for particle identification.

The expected track lengths in the HIBEAM TPC are approximately \(10~\mathrm{cm}\)~\cite{Santoro:2024lvc}, compared to about \(2~\mathrm{m}\) in the ALICE TPC~\cite{alme2010alice}. The TPC surrounds a central vacuum tube containing the thin carbon-foil target. Charged particles produced in an annihilation exit the tube and are subsequently tracked in the surrounding gas volume. Multiple scattering in the aluminium wall of the vacuum tube can limit the pointing accuracy when extrapolating tracks back to the annihilation vertex. In the later NNBAR stage, the tracking volume is expected to be larger, but the reconstruction of short, straight track segments remains relevant for prototype development.

These challenges are addressed in this work by studying a TPC prototype for the HIBEAM detector concept. The prototype matches the short track lengths relevant for HIBEAM and uses a zigzagpad grid to improve charge sharing in the readout plane. A reconstruction toolkit~\cite{ivanov2006track} for short tracks has been developed and validated with simulated data, cosmic measurements and proton-elastic-scattering measurements. Using these datasets, key operating parameters of the TPC prototype are studied, including the drift field, shaping time and GEM voltages. In addition, tracking-performance metrics, residual distributions and reconstructed \(dE/dx\) distributions are investigated, with \(dE/dx\) used here as a quantity proportional to the deposited energy per unit length. The results provide guidance for the full-scale HIBEAM TPC prototype, with possible relevance for later NNBAR tracking studies.

This article is organised as follows. First, the role of the TPC in the annihilation detector is summarised. The HIBEAM TPC concept and the TPC prototype are then introduced, followed by the reconstruction algorithm development. The experimental datasets, obtained from cosmic-trigger and CCB proton-trigger measurements, are described, and the reconstruction procedure is validated with simulated and experimental data. Finally, the implications for the future HIBEAM TPC design and possible next steps in prototype development are discussed.

\section{TPC Role in the Annihilation Detector}
\label{sec:tpc_role}

The signature of neutron--antineutron oscillation is the annihilation of an antineutron on a carbon target surrounded by an annihilation detector. Such interactions produce a multi-pion state, typically including charged pions and photons from neutral-pion decays, with particle momenta of order \(100\)--\(300~\mathrm{MeV}\)~\cite{ann1,ann2,ann3}. The branching ratios for free \(\bar{n}\)--nucleon annihilation show that most final states contain charged pions and neutral pions~\cite{Abe2015nnbar}. This motivates an annihilation-detector concept combining charged-particle tracking and electromagnetic calorimetry.

Annihilation in a nucleus, carbon in this case, opens additional ways to dissipate energy and momentum by re-scattering in the nucleus. Neutrons, protons and the recoiling remnant nucleus can carry away part of the energy and momentum. Consequently, the requirements on energy and momentum conservation cannot be as strict in \(\bar{n}\)--C annihilation as in the free \(n\)--\(\bar{n}\) case. Charged-particle tracking is therefore needed to reconstruct the event topology and to determine whether the charged-particle topology is consistent with an annihilation signal or with background particles entering the detector from outside.

\begin{figure}[ht!]
\centering
\includegraphics[width=0.8\textwidth]{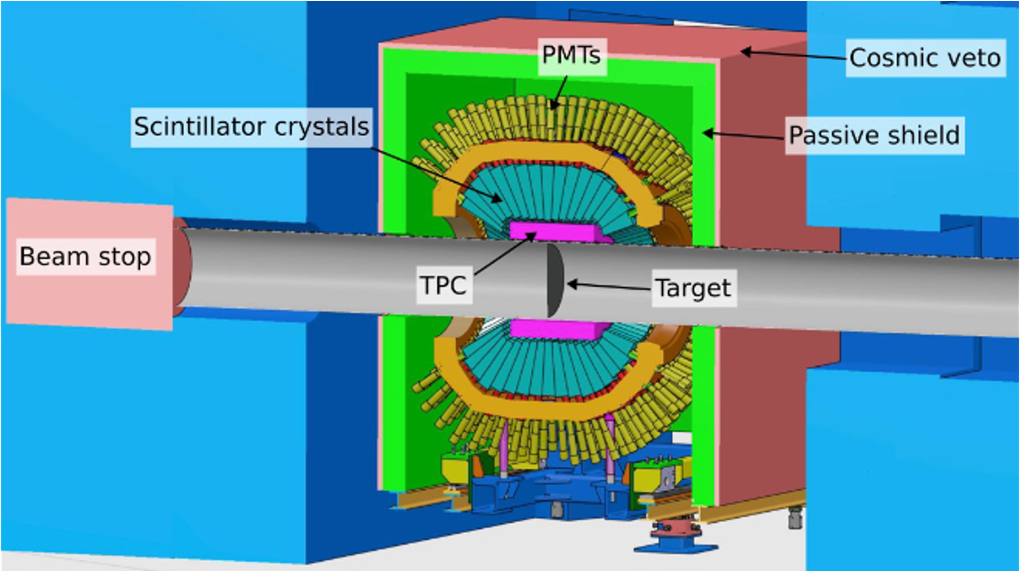}
\caption{Overview of the HIBEAM annihilation-detector concept. The TPC surrounds the vacuum tube and carbon target foil and provides charged-particle tracking. Adapted from Ref.~\cite{persson2024design}.}
\label{fig:tpc_linus}
\end{figure}

A schematic of the HIBEAM annihilation detector is shown in Fig.~\ref{fig:tpc_linus}. Incoming antineutrons travelling from the right in the vacuum tube annihilate in a thin carbon foil. The focus of this work is the TPC, which surrounds the target region and provides charged-particle tracking.

Reconstructed TPC tracks are extrapolated towards the annihilation target region to distinguish annihilation-like topologies from background particles entering the detector from outside. The track directions also allow the TPC information to be matched to energy deposits in the surrounding calorimeter. In addition, the TPC measures the energy loss, \(dE/dx\), which can provide complementary information for distinguishing charged-pion signal candidates from background particles.

For the full detector, the TPC must provide three-dimensional track information. The coordinates in the readout plane are reconstructed from the segmented charge readout, while the drift coordinate is obtained from the signal arrival time. An external event-time reference is required to convert this time into an absolute drift position. In HIBEAM, this time reference is expected to be provided by the surrounding detector systems.

\subsection{HIBEAM TPC Concept}
\label{sec:hibeam_tpc_concept}

The TPC at HIBEAM is designed as a cylindrical gas chamber that surrounds the thin carbon target and follows the beam axis~\cite{santoro2025hibeam,Hehl2025_BSc}. The cylinder has inner and outer diameters of about \(40~\mathrm{cm}\) and \(60~\mathrm{cm}\), respectively, and an active length of order \(50\)--\(60~\mathrm{cm}\)~\cite{Hehl2025_BSc}. The TPC cylinder is divided into two drift regions by a central cathode plane held at approximately \(-15~\mathrm{kV}\). This results in a maximum drift distance of about \(25~\mathrm{cm}\) from the cathode to the GEM-based readout plane on each side.

The TPC measures charged-particle tracks by recording the ionisation they produce in a gas volume. As a particle traverses the chamber, it creates electron--ion pairs along its path. In the presence of the drift electric field, the liberated electrons drift with approximately constant velocity towards the readout plane. When the electron cloud reaches the amplification stage, it is multiplied in the GEM stack via avalanche amplification, producing a measurable charge signal. The amplified charge is collected on a segmented pad plane and digitised by the front-end electronics as time-sampled charge pulses. The charge distribution over the pads provides the \(x\) and \(y\) coordinates in the readout plane, while the signal arrival time is converted to the drift coordinate using the calibrated drift velocity once the event time is known.

The anode planes are located at the two endcaps and are held close to ground potential. Avalanche multiplication is provided by a stack of GEMs mounted in front of each readout plane. A quadruple GEM configuration has been chosen to reach the required gas gain while keeping the ion backflow into the drift volume under control~\cite{Sauli1997GEM,Hehl2025_BSc}. Dividing the chamber into two drift regions reduces the maximum drift distance, which helps to limit diffusion and improve the spatial resolution. It also lowers the required cathode voltage for a given drift field. The TPC is planned to operate with an Ar/CO\(_2\) 80/20\,\% gas mixture. Two readout planes consists of 3200 pads, with one electronic channel per pad~\cite{santoro2025hibeam}.

The absence of a magnetic field also affects the optimisation of the HIBEAM TPC. In many TPC applications, a magnetic field reduces transverse diffusion during the drift, but this is not available in HIBEAM. At the same time, the HIBEAM TPC is compact and has a short maximum drift distance, so it is expected to operate with relatively small diffusion. This is especially relevant for tracks close to the GEM readout plane, where diffusion is minimal. The readout geometry and operating conditions must therefore provide sufficient position-sensitive charge sharing even for narrow charge clouds.

The readout plane of this compact TPC concept is segmented into zigzag pads,
which provide position-sensitive charge sharing between neighbouring pads even when the electron cloud reaching the plane is narrow. The small transverse diffusion expected inside the drift volume of TPC helps to keep nearby tracks separable, but would provide only limited charge sharing for an equivalent rectangular pad geometry. Therefore, the zigzag structure makes it possible to reconstruct centroid positions from the charge distribution over neighbouring pads, rather than relying on single-pad signals. The charge sharing capability of the zigzag shaped readout is demonstrated in Figure~\ref{fig:pad_columns_comparison}.

Another consequence of the compact geometry is that many relevant tracks are short, with a projected length of order \(10~\mathrm{cm}\), and some tracks may be close to the drift direction. The reconstruction must therefore remain robust with a limited number of pad rows and with time profiles that can be broadened by the track inclination in the drift plane. Since multiple scattering before the TPC can limit the achievable back-pointing accuracy, the reconstruction studies focus on robust short-track reconstruction rather than on pursuing extreme intrinsic spatial resolution.

The prototype studied in this work originates from the ILC TPC development programme~\cite{Shoji2018LCTPC} and was further developed at Lund University. Significant work was also carried out at Lund University to adapt and further develop the readout electronics and data-acquisition chain~\cite{Hehl2025_BSc}. Although the prototype uses a simpler rectangular field-cage geometry than the final cylindrical HIBEAM concept, its active dimensions closely match the relevant HIBEAM TPC scale, with track lengths down to \(10~\mathrm{cm}\) and a maximum drift length of \(23.1~\mathrm{cm}\). It is therefore well suited for studying short-track reconstruction, charge sharing with zigzagpads and operating parameters relevant for the HIBEAM TPC. The prototype geometry, readout and operating parameters are described in the following section.

\subsubsection{Readout Electronics}
\label{sec:readout_electronics}

The prototype readout plane is connected to two front-end Multi Chip Modules
(MCMs), each hosting eight SALTRO16 chips with 16 channels per chip. This
gives 256 connected readout channels in total. The SALTRO16 chip combines
preamplification, pulse shaping, analogue-to-digital conversion and a memory pipeline~\cite{aspell2013saltro16,Gaspari2012SoC}. The signals are sampled
at \(20~\mathrm{MHz}\), corresponding to \(50~\mathrm{ns}\) per timestamp. After preamplification, the signal is shaped before digitisation. The shaping time affects the temporal spread of the digitised ADC profile: shorter shaping
preserves more of the time structure of the arriving charge, while longer shaping produces a smoother and wider response.

\begin{figure}[tb]
  \centering
  \includegraphics[width=0.95\textwidth]{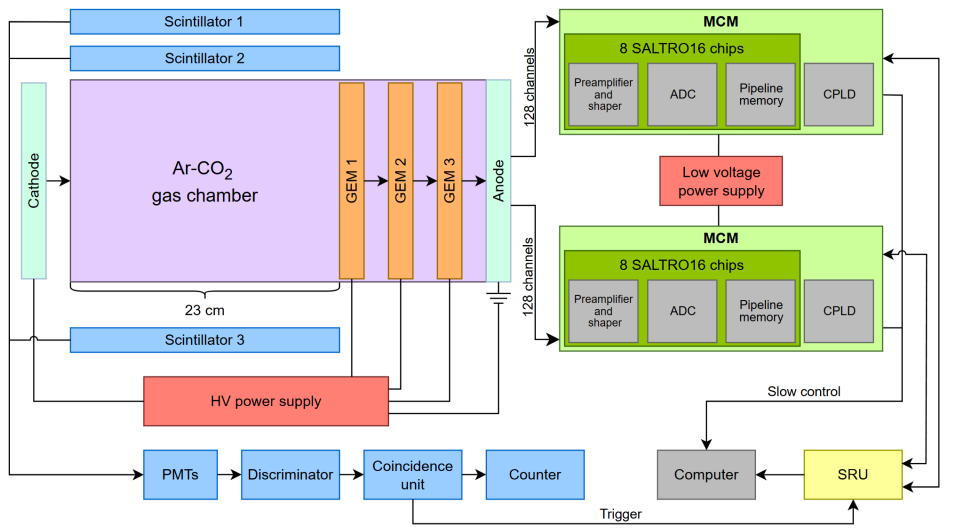}
  \caption{Schematic overview of the prototype readout and data-acquisition
  system, including the TPC, scintillator trigger system, front-end MCMs and
  SRU. Adapted from Ref.~\cite{Hehl2025_BSc}.}
  \label{fig:tpc_readout_daq}
\end{figure}

The front-end MCMs and data-acquisition system were developed by the Lund
University group for the Linear Collider TPC (LCTPC) collaboration within the EU project AIDA. The system was constructed with funding from the Lund University Faculty of Science infrastructure programme. The MCM output is serialised by a Complex Programmable Logic Device (CPLD) and transferred to the Scalable Readout Unit (SRU), which provides clock distribution and data
transfer to the acquisition computer~\cite{Hehl2025_BSc,Martoiu2013SRS}. The readout and data-acquisition system is shown schematically in
Figure~\ref{fig:tpc_readout_daq}.

\FloatBarrier
\section{Experimental Tests with TPC Prototype}
\subsection{Cosmic Measurements}
To validate the tracking procedure and characterise the TPC prototype, cosmic measurements were used as a natural and continuous source of charged particles. At sea level, this sample is expected to be dominated by atmospheric muons, many of which have GeV-scale energies and behave approximately as minimum-ionising particles in the Ar/CO$_2$ gas mixture. For such muons, the energy loss is close to the broad minimum of the Bethe--Bloch curve and changes only slowly with energy~\cite{PDG2022,PassageMatter}. This makes cosmic-ray measurements a convenient reference sample for detector studies.

The cosmic measurements were performed to study how the detector performance depends on the main operating parameters and on the track geometry. The chamber was continuously flushed with an Ar/CO$_2$ (80/20) gas mixture at a flow rate of about 12 L/h, and the gas flow was started one day before data taking to limit oxygen and water contamination. The signal was acquired following the muons triggering plastic scintillators mounted above and below the TPC. A coincidence of the scintillator signals defines the trigger and the start time for the drift. Data were collected for different settings of the drift field (243–500 V/cm), GEM voltages (approximately 340–360 V per GEM) and shaping times (60 and 120 ns). For all measurements analysed here, a three-scintillator trigger was used (two scintillators above the chamber and one below) to suppress random coincidences. \cite{Hehl2025_BSc}
\subsection{Elastic Proton-Scattering Measurements at CCB}
\label{sec:ccb_proton_test}

\begin{figure}[tb]
    \centering
    \includegraphics[width=0.72\textwidth]{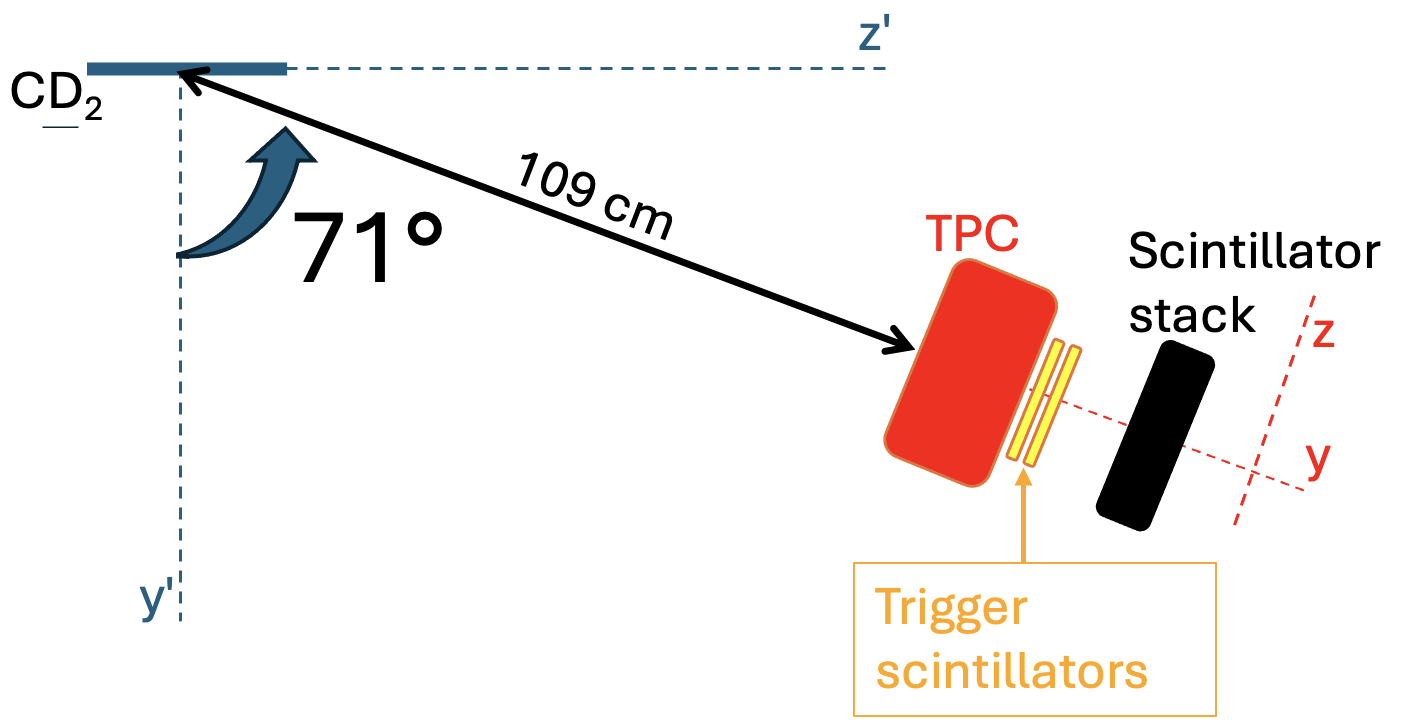}
    \caption{Schematic illustration of the target--TPC geometry in the CCB experiment for the baseline TPC orientation. The TPC arm was placed at \(71^\circ\) with respect to the proton beam direction. Trigger scintillators located behind the TPC defined the geometrical acceptance used to identify trigger tracks in the reconstructed data.}
    \label{fig:ccb_target_tpc_sketch}
\end{figure}

In addition to the cosmic measurements, the TPC prototype was tested in an in-beam experiment at the Cyclotron Centre Bronowice (CCB) in Kraków, Poland. In this measurement, a \(190~\mathrm{MeV}\) proton beam was directed onto a CD$_2$ target, where elastic-scattering reactions could occur. The TPC arm was placed at \(71^\circ\) with respect to the beam direction and a second scintillator arm was placed at \(38^\circ\). Coincidences between the two arms were used to select candidate elastic-scattering events. In the TPC arm, trigger scintillators located behind the chamber defined the geometrical acceptance used to identify trigger tracks in the reconstructed data. Reaching the trigger scintillator therefore imposed an approximate range-based lower-energy selection on the trigger-track sample. For a \(1~\mathrm{cm}\) plastic scintillator, the kinetic energy required for a proton to traverse the scintillator is about \(31~\mathrm{MeV}\), while the corresponding estimate for a deuteron is about \(62~\mathrm{MeV}\), using the same energy-per-nucleon argument. The baseline geometry of the CCB measurement is shown in Figure~\ref{fig:ccb_target_tpc_sketch}.

The CCB data provide an important complement to the cosmic measurements, especially for studying tracks inclined in the drift plane. The cosmic measurements are expected to be dominated by atmospheric muons, which behave approximately as minimum-ionising particles in the Ar/CO$_2$ gas mixture. In contrast, the particles crossing the TPC in the CCB trigger-track sample are protons with a much lower velocity. Assuming proton kinetic energies of about \(30\)--\(40~\mathrm{MeV}\) at the TPC position gives
\[
\beta =
\sqrt{1-\frac{1}{\gamma^2}},
\qquad
\gamma = 1 + \frac{T_p}{m_pc^2},
\]
which corresponds to \(\beta \approx 0.25\)--\(0.28\). In this velocity range, the ionisation term in the Bethe--Bloch formula is dominated by the approximate \(1/\beta^2\) dependence. This gives an ionisation scale of about \(12\)--\(16\) times larger than for relativistic particles with \(\beta \approx 1\), and about \(12\)--\(13\) for protons close to \(40~\mathrm{MeV}\).

\section{Track Reconstruction of TPC Events}
\label{sec:track_reconstruction}

The tracking algorithm reconstructs charged-particle tracks in TPC events from digitised ADC values recorded for individual pad columns and timestamps in each pad row. Within each pad row, neighbouring ADC values are grouped into clusters and converted into three-dimensional centroids. Track candidates are then built by linking compatible centroids from adjacent pad rows.

The procedure is intended for the short, approximately straight tracks recorded
with the prototype TPC. It is applied to both low-multiplicity and
higher-multiplicity experimental data, and therefore has to handle both
single-track and multi-track events. Frequent geometrical crossings are not
expected for the majority of the track topologies considered here.

The reconstruction proceeds in three main steps:
\begin{enumerate}
    \item clusters of neighbouring ADC values are identified within each pad row,
    using neighbouring pad columns and timestamps,
    \item each cluster is converted into a centroid, representing a reconstructed
    space point with coordinates \((x,y,z)\) calculated from the ADC values
    belonging to that cluster. The centroid also carries an integrated ADC value
    and coordinate uncertainties,
    \item these centroids are linked across pad rows and fitted with straight
    lines in the \(yx\) and \(yz\) projections.
\end{enumerate}

\begin{figure}[H]
  \centering
  \begin{subfigure}{0.48\textwidth}
    \centering
    \includegraphics[width=\textwidth]{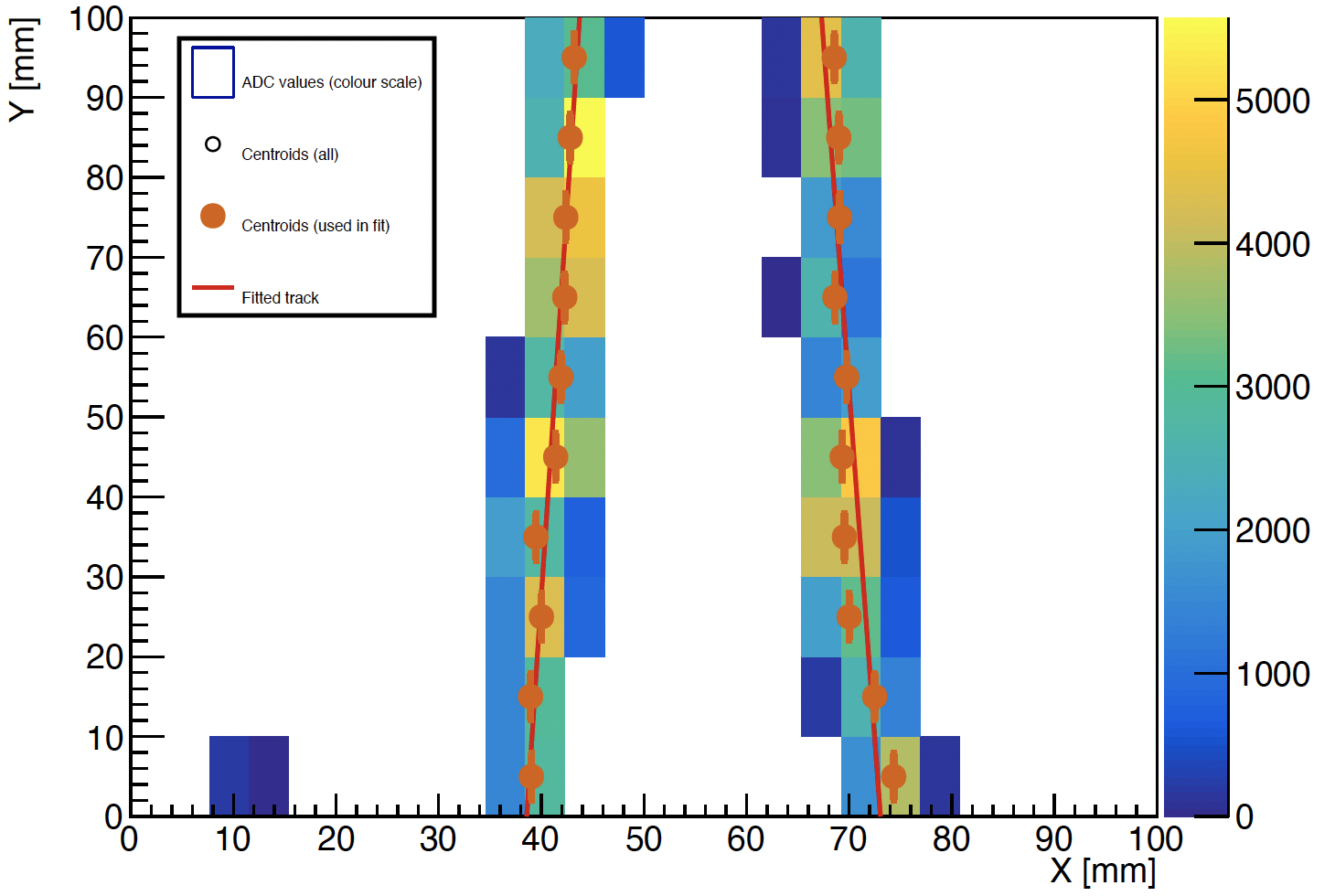}
    \caption{\(yx\) projection.}
  \end{subfigure}
  \hfill
  \begin{subfigure}{0.49\textwidth}
    \centering
    \includegraphics[width=\textwidth]{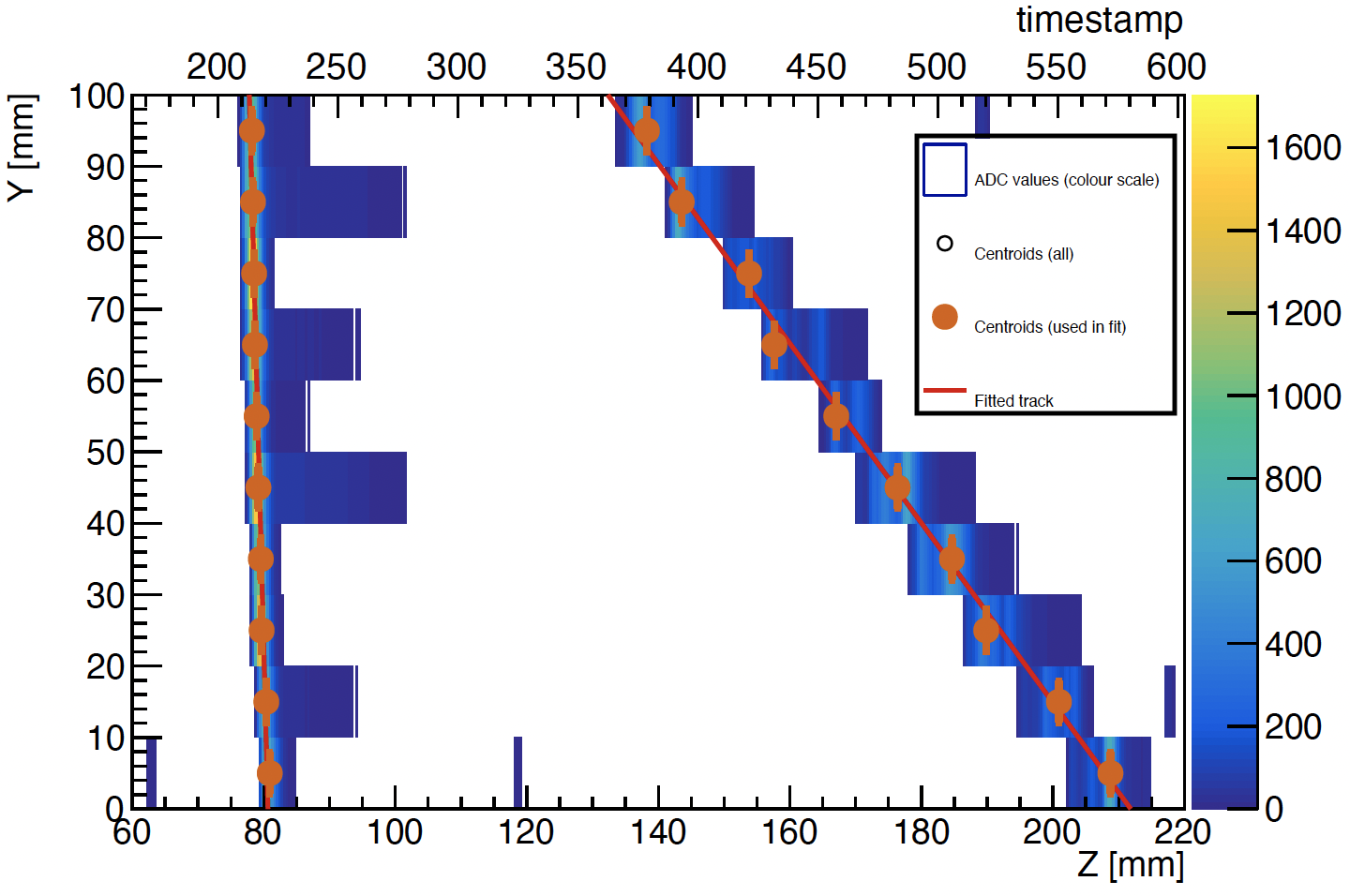}
    \caption{\(yz\) projection.}
  \end{subfigure}
  \caption{Example of a reconstructed double-track cosmic event in the two fitted projections.}
  \label{fig:reco_ccb_multitrack}
\end{figure}

\begin{figure}[H]
  \centering
  \includegraphics[width=0.70\textwidth]{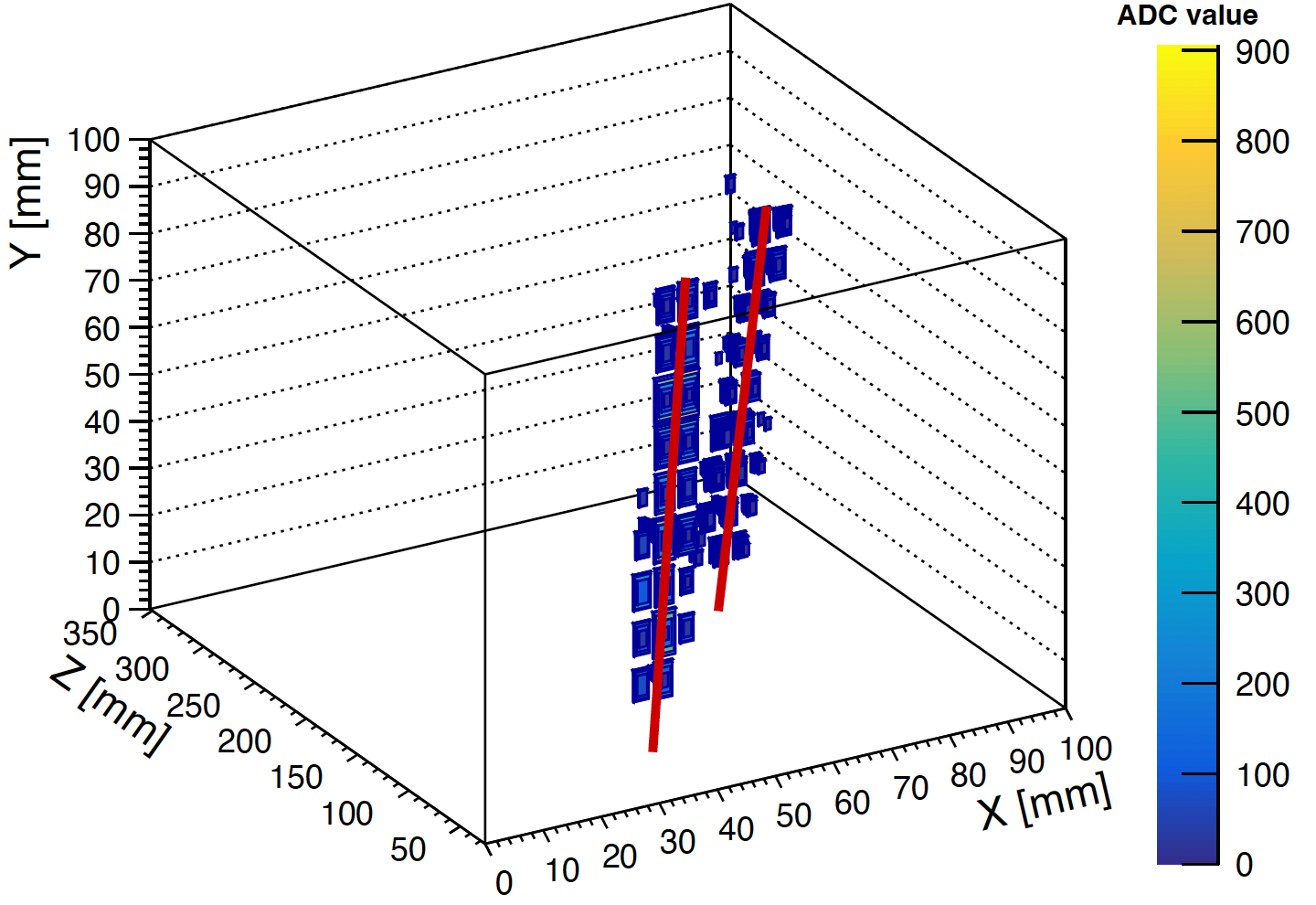}
  \caption{Three-dimensional view of the same reconstructed double-track cosmic event as in Figure~\ref{fig:reco_ccb_multitrack}.}
  \label{fig:reco_ccb_multitrack_3d}
\end{figure}

An example of a reconstructed experimental cosmic event is shown in
Figures~\ref{fig:reco_ccb_multitrack} and~\ref{fig:reco_ccb_multitrack_3d}.
The event contains two tracks and illustrates the type of multi-track topology
that is relevant for the present reconstruction study.

\subsection{Cluster Finding in Individual Pad Rows}

The first reconstruction step groups neighbouring ADC values into clusters
within each pad row. The clustering is performed in the two-dimensional space spanned by pad column and timestamp, so that the time structure of the ADC
distribution is retained from the beginning of the reconstruction. The output of this step is a set of pad-row clusters, each corresponding to a localised ionisation deposit in one row of the readout plane.

\begin{figure}[H]
  \centering
  \includegraphics[width=0.92\textwidth]{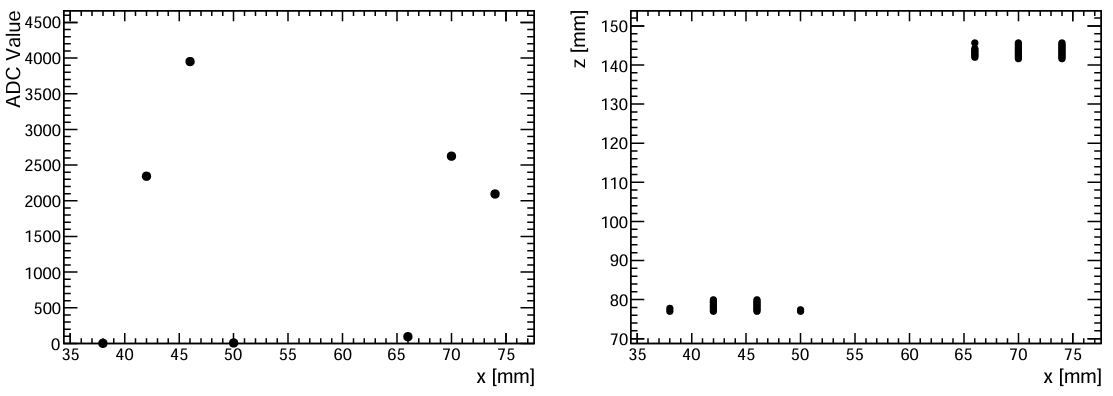}
  \caption{Example of two clusters reconstructed in the pad row centred at
\(y=85~\mathrm{mm}\) of the double-track cosmic event shown in
Figures~\ref{fig:reco_ccb_multitrack} and~\ref{fig:reco_ccb_multitrack_3d}.
The ADC values are integrated over the whole cluster. The ADC values are integrated over the whole cluster.}
  \label{fig:cluster_span_example}
\end{figure}

Figure~\ref{fig:cluster_span_example} shows two clusters from one pad row of
the same double-track cosmic event shown in Figures~\ref{fig:reco_ccb_multitrack}
and~\ref{fig:reco_ccb_multitrack_3d}. The ADC values are integrated over the
whole cluster. The figure shows both the cluster span in \(x\), through the
integrated ADC value in each pad column, and the cluster span in \(z\), through
the ADC-sample positions in the drift direction.

For more complicated topologies, the cluster finder also includes an
additional pulse-separation logic. After a pulse maximum is identified, the falling tail of the ADC profile is followed. In this region, an exponential
moving average (EMA) is used to estimate the tail level smoothly. If the ADC
profile subsequently shows a sustained rise above this tail level, the
algorithm interprets this as the onset of a second cluster and splits the
cluster. In plain terms, this allows partially overlapping ADC time profiles within one pad row to be separated. This treatment is mainly relevant for
more ambiguous cases and for reconstruction-development studies, while the majority of the expected track topologies are already well described by the
basic clustering procedure.

\subsection{Centroid Reconstruction}

Once a cluster has been identified, it is represented by a single
three-dimensional centroid. Let \(q_i\) denote the ADC value of entry \(i\), with pad-column index \(c_i\) and timestamp \(t_i\). The \(x\) coordinate is
obtained from the ADC-weighted mean pad-column position,
\begin{equation}
x_c
=
\Delta x_{\mathrm{pad}}
\left(
\frac{1}{2}
+
\frac{\sum_i q_i c_i}{\sum_i q_i}
\right),
\label{eq:centroid_x}
\end{equation}
where \(\Delta x_{\mathrm{pad}}\) is the pad-column pitch. This charge sharing between neighbouring pad columns is particularly important for the zig-zag
readout geometry, where the position information is carried by the relative
distribution of ADC values over neighbouring pads.

The \(y\) coordinate is taken as the centre of the pad row,
\begin{equation}
y_c
=
\Delta y_{\mathrm{row}}
\left(
r + \frac{1}{2}
\right),
\label{eq:centroid_y}
\end{equation}
where \(r\) is the pad-row index and \(\Delta y_{\mathrm{row}}\) is the pad-row
height. Since the centroid is assigned to the middle of the row, the
uncertainty in \(y\) is taken from the width of a uniform distribution across
the pad row,
\begin{equation}
\sigma_y = \frac{\Delta y_{\mathrm{row}}}{\sqrt{12}}.
\label{eq:sigma_y}
\end{equation}

For the drift coordinate, the ADC values in the cluster are first summed over
pad columns to form a pulse profile as a function of timestamp. Around the
pulse maximum, an asymmetric time window is defined. The window follows the typical pulse shape, with a faster rise and a slower falling tail. The
centroid timestamp is then calculated as the ADC-weighted mean timestamp
inside this window,
\begin{equation}
t_c
=
\frac{\sum\limits_{t=t_{\min}}^{t_{\max}} t\,Q(t)}
     {\sum\limits_{t=t_{\min}}^{t_{\max}} Q(t)},
\label{eq:centroid_t}
\end{equation}
where \(Q(t)\) is the summed ADC value at timestamp \(t\). The corresponding
integrated ADC value of the centroid is
\begin{equation}
Q_c = \sum\limits_{t=t_{\min}}^{t_{\max}} Q(t).
\label{eq:centroid_charge}
\end{equation}

\begin{figure}[ht!]
  \centering
  \begin{subfigure}{0.49\textwidth}
    \centering
    \includegraphics[width=\textwidth]{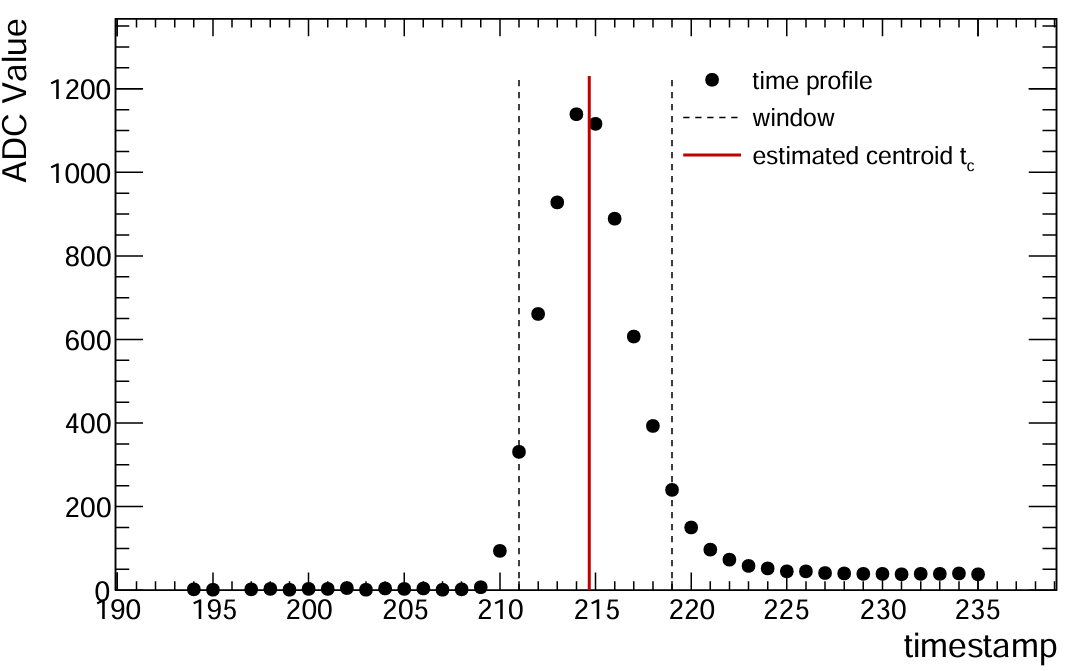}
\caption{Cluster associated with the track at smaller \(x\) at \(y=85~\mathrm{mm}\).}
\end{subfigure}
  \hfill
  \begin{subfigure}{0.49\textwidth}
    \centering
    \includegraphics[width=\textwidth]{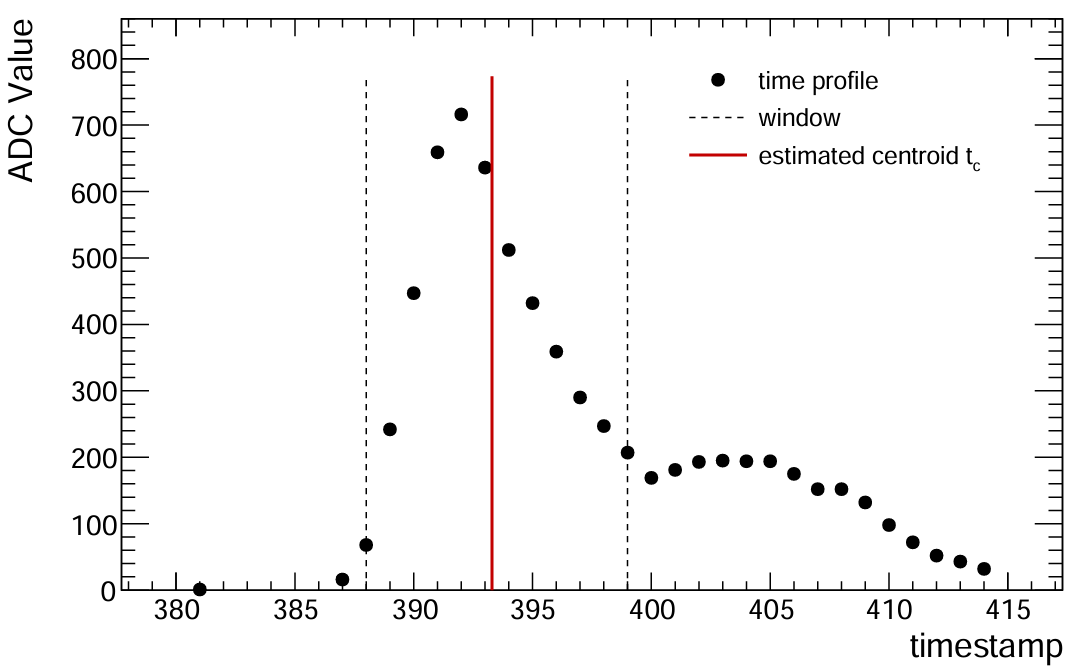}
\caption{Cluster associated with the track at larger \(x\) at \(y=85~\mathrm{mm}\).}
\end{subfigure}
  \caption{Time profiles for the two clusters shown in Figure~\ref{fig:cluster_span_example}. The ADC values are summed over the whole cluster.}
  \label{fig:cluster_time_profile_example}
\end{figure}

Figure~\ref{fig:cluster_time_profile_example} shows the corresponding time
profiles for the two clusters shown in Figure~\ref{fig:cluster_span_example}.
The dashed lines indicate the asymmetric window around the pulse maximum, and
the red line shows the ADC-weighted centroid timestamp \(t_c\) calculated
within this window.

The centroid timestamp is converted to the drift coordinate using the known
drift velocity. Since one timestamp corresponds to \(50~\mathrm{ns}\), the
drift coordinate is calculated as
\begin{equation}
z_c = t_c \cdot 50~\mathrm{ns} \cdot v_{\mathrm{drift}}.
\label{eq:centroid_z}
\end{equation}
Equivalently, this can be written as
\begin{equation}
z_c = \alpha t_c,
\label{eq:centroid_z_alpha}
\end{equation}
where \(\alpha\) is the timestamp-to-distance conversion factor used in the
reconstruction.

The coordinate uncertainties in \(x\) and \(z\) are estimated from the spread
of the ADC distribution around the centroid. In the pad direction,
\begin{equation}
\sigma_x
=
\frac{\sqrt{\sum_i q_i^2 (x_i-x_c)^2}}{\sum_i q_i},
\label{eq:sigma_x}
\end{equation}
where \(x_i\) is the position corresponding to pad column \(c_i\). In the
drift direction,
\begin{equation}
\sigma_z
=
\alpha\,
\frac{\sqrt{\sum\limits_{t=t_{\min}}^{t_{\max}} Q(t)^2 (t-t_c)^2}}
     {\sum\limits_{t=t_{\min}}^{t_{\max}} Q(t)}.
\label{eq:sigma_z}
\end{equation}
If all ADC values are concentrated in a single pad column or a single
timestamp, the corresponding uncertainty is limited from below by the width
of one pad column or one timestamp. The lower limits are therefore
\(\Delta x_{\mathrm{pad}}/\sqrt{12}\) in \(x\) and
\(\alpha/\sqrt{12}\) in \(z\).

\subsection{Track Building and Line Fitting}

After the centroid reconstruction, the centroids are linked across pad rows to form track candidates. Candidate links are considered only if they satisfy geometrical limits in the pad-row, pad-column and drift directions. For two candidate centroids \(a\) and \(b\), let
\[
  \Delta r = |r_b-r_a|,
\]
where \(r_a\) and \(r_b\) are the pad-row indices of the two centroids. The
link is accepted only if
\begin{align}
1 \leq \Delta r \leq N_{\mathrm{row}}^{\max}, \\
|x_b-x_a| \leq \Delta r\,\Delta x_{\max}, \\
|z_b-z_a| \leq \Delta z_{\max}.
\end{align}
Here \(N_{\mathrm{row}}^{\max}\) is the maximum allowed pad-row jump,
\(\Delta x_{\max}\) is the maximum allowed change in \(x\) for one pad-row
step, and \(\Delta z_{\max}\) is the maximum allowed change in the drift
coordinate. In the reconstruction used here, links between neighbouring pad rows are allowed, and one missing pad row can also be skipped if the same
geometrical limits are satisfied. This helps to keep a track continuous when one pad row contains an outlying centroid or no usable cluster.

For the accepted candidate track links, the basic geometrical cost is taken
as the squared Euclidean separation,
\begin{equation}
c^{\mathrm{geom}}_{ab}
=
(x_a-x_b)^2 + (y_a-y_b)^2 + (z_a-z_b)^2 .
\end{equation}
This is sufficient for the majority of the expected track topologies, where
only one track per event is expected.

When more than one candidate link is present, the matching is formulated as
a one-to-one assignment problem. The allowed links and their costs define a
cost matrix, and the assignment is solved with the Hungarian algorithm~\cite{Kuhn1955Hungarian}.
In ordinary single-track cases, or in events where only one compatible
continuation is available, this gives the same result as a nearest-neighbour
choice. The advantage of this procedure appears when several centroids in one
pad row can be linked to several centroids in the next allowed pad rows,
including the case where one pad row is skipped. A local nearest-neighbour
choice treats each link separately. Therefore, it can choose a combination of
links between centroids that is not optimal for the candidate set as a whole.
The Hungarian assignment instead selects the one-to-one set of links with the
lowest total cost among the allowed candidates.

For more challenging topologies, such as crossing tracks where the clusters
from two tracks overlap in one or several pad rows, additional terms can be included in the cost function. These terms are not central to the reconstruction
of the experimental data discussed in this work, where the majority of the
expected track topologies are reconstructed using the Euclidean-distance cost.
They were used mainly as safeguards for difficult multi-track cases: a
continuity term can prefer candidates that follow the local track direction,
while a mild length bias can favour extending an already established track
rather than breaking it because of an outlier or a distorted centroid close to a crossing point.

The final track candidates are fitted with straight lines in the \(yx\) and \(yz\)
projections, using \(y\) as the independent coordinate:
\begin{align}
x(y) &= a_{yx} + b_{yx} y, \\
z(y) &= a_{yz} + b_{yz} y.
\end{align}
given the assumption of the straight tracks in the HIBEAM TPC. The fitted tracks are then used in the subsequent analysis of residuals, track quality and \(dE/dx\)-related observables. This tracking procedure was validated on both the experimental and simulated data as reported in Section~\ref{sec:validation_application}.

\FloatBarrier
\section{Validation Methodology and Summary of the Reconstruction Performance on Simulated and Experimental Data}
\label{sec:validation_application}

\subsection{Defining the Residual Widths}

The residual widths reported in the following tables were obtained using
leave-one-out fits. Each centroid was removed in turn, the track was refitted,
and the residual was calculated as the difference between the removed centroid
and the refitted track prediction at the same pad-row position. The quantities
\(\sigma^{\mathrm{LOO}}_x\) and \(\sigma^{\mathrm{LOO}}_z\) denote the fitted
Gaussian widths of the residual distributions in the \(x\) and \(z\)
coordinates.

\subsection{Defining the Tracking Efficiency}

The tracking efficiency was defined as the fraction of non-accidental-coincidence
events with at least one good reconstructed track. Events with a summed centroid ADC below 1000 counts or with fewer than four reconstructed centroids were classified as accidental-coincidence candidates and were excluded from the
efficiency denominator. A good reconstructed track was required to satisfy a
reduced-\(\chi^2\) selection in both the \(yx\) and \(yz\) projections.

For the reference cosmic measurement at 500 V/cm, tightening the reduced-\(\chi^2\)
requirement from \(\chi^2_\nu<10\) to \(\chi^2_\nu<2.5\) produced only a modest
loss of statistics while slightly improving the residual widths. A stricter
requirement of \(\chi^2_\nu<1\) removed a much larger fraction of the tracks.
The requirement \(\chi^2_\nu<2.5\) was therefore chosen as the standard
good-track definition for the main reconstruction-performance summary.

The simulation data allow the reconstruction algorithm to be tested under
controlled conditions, with known track geometry. The perpendicular simulation
data provide a simple reconstruction case, for which the standard
tracking-efficiency definition can be applied directly. The crossing simulation
data contain crossing tracks by construction and are used only as a stress test
of the centroid-linking step. Multi-track events are expected in the CCB data
and in the future search for \(n \rightarrow \bar{n}\) at HIBEAM, but frequent geometrical crossings between tracks are not expected. Therefore, the crossing simulation data are  not
included in the main tracking-efficiency comparison and are discussed
separately in Section~\ref{sec:crossing_stress_test}.

\begin{table}[ht!]
\centering
\small
\renewcommand{\arraystretch}{1.15}
\setlength{\tabcolsep}{4pt}
\caption{Study of the reduced-\(\chi^2\) threshold used to define good reconstructed tracks.}
\label{tab:chi2_cut_choice}
\begin{tabular}{llcccc}
\hline
Data & \(\chi^2_\nu\) cut & Tracks kept & Retained [\%] &
\(\sigma^{\mathrm{LOO}}_x\) [mm] &
\(\sigma^{\mathrm{LOO}}_z\) [mm] \\
\hline
Cosmic, 500 V/cm & \(<10.0\) & 435 & 100.0 & 0.5126 & 0.4068 \\
Cosmic, 500 V/cm & \(<5.0\)  & 427 & 98.2  & 0.5097 & 0.4050 \\
Cosmic, 500 V/cm & \(<2.5\)  & 415 & 95.4  & 0.5062 & 0.3985 \\
Cosmic, 500 V/cm & \(<1.0\)  & 326 & 74.9  & 0.5113 & 0.3676 \\
\hline
\end{tabular}
\end{table}

\subsection{Defining the Energy Loss \(dE/dx\)}
The energy loss \(dE/dx\) was calculated from the ADC values assigned to the reconstructed centroids. For each centroid, the integrated ADC value was normalised by the corresponding track-length interval, giving a relative energy-loss quantity in ADC/cm. This quantity is expected to scale with the physical \(dE/dx\), but it should not be interpreted as an absolute energy-loss measurement. Track-level \(dE/dx\) values were then obtained from the centroid \(dE/dx\) values assigned to each reconstructed track.

\subsection{Crossing Simulation Data as a Stress Test}
\label{sec:crossing_stress_test}

The crossing simulation data contain two crossing tracks per event and were
used to test the track-linking algorithm in a deliberately challenging crossing
topology. This provides a stress test of the reconstruction algorithm. For this
stress-test metric, a track was counted as reconstructed if it satisfied the
looser requirement \(\chi^2_\nu<20\).

With this requirement, the average number of reconstructed tracks per event is \(1.70\). The reconstruction becomes less reliable when the two crossing tracks are close
in the drift projection. The reconstruction becomes less reliable when the two crossing tracks are close
in the drift projection. Events in which both reconstructed tracks satisfy the \(\chi^2_\nu<20\) requirement have a median relative angle of \(30.2^\circ\) in the \(yz\) projection, while events where at least one track fails this requirement have a median relative angle of \(13.2^\circ\). A visual inspection indicated that approximately \(10~\%\) of the crossing
simulation data correspond to cases with strong geometrical overlap between the
two tracks, for which the two trajectories are not clearly separable by eye. These cases are one source of missed double-track reconstruction in the crossing simulation data.

\subsection{Validation and Application of the Track Reconstruction Algorithm}

\begin{table}[H]
\centering
\small
\renewcommand{\arraystretch}{1.15}
\setlength{\tabcolsep}{3pt}
\caption{Summary of track reconstruction performance for the perpendicular
simulation data and the experimental data. The tracking efficiency is defined
as the fraction of non-accidental-coincidence events with at least one good
reconstructed track, using \(\chi^2_\nu<2.5\) in both projections. The crossing
simulation data are treated separately as a stress test in
Section~\ref{sec:crossing_stress_test}.}
\label{tab:tpc_main}
\begin{tabularx}{\textwidth}{>{\raggedright\arraybackslash}p{3.5cm}CCCCCC}
\hline
Data &
\shortstack{Tracking\\eff. [\%]} &
\shortstack{Good tracks\\per event} &
\shortstack{\(\sigma^{\mathrm{LOO}}_x\)\\{[}mm{]}} &
\shortstack{\(\sigma^{\mathrm{LOO}}_z\)\\{[}mm{]}} &
\shortstack{median\\\(\chi^2_{\nu,yx}\)} &
\shortstack{median\\\(\chi^2_{\nu,yz}\)} \\
\hline
Simulation, perpendicular
& 97.90 & 0.99 & 0.063 & 0.271 & 0.018 & 0.290 \\
\hline
Cosmic 500 V/cm
& 96.66 & 1.08 & 0.519 & 0.441 & 0.564 & 0.318 \\

Cosmic 243 V/cm
& 96.60 & 1.04 & 0.537 & 0.310 & 0.882 & 0.474 \\

CCB perpendicular
& 96.87 & 1.53 & 0.457 & 0.283 & 0.583 & 0.371 \\

CCB tilted \(45^\circ\)
& 56.86 & 0.66 & 0.536 & 1.304 & 1.530 & 0.195 \\

CCB perpendicular higher-multiplicity
& 99.89 & 5.17 & 0.581 & 0.574 & 0.851 & 1.178 \\
\hline
\end{tabularx}
\end{table}

The first row of Table~\ref{tab:tpc_main} shows the perpendicular simulation data, which represent the simplest reconstruction case in the comparison. As expected, this configuration gives the smallest residual widths, since the tracks are fully perpendicular in both the transverse \(yx\) and drift \(yz\) projections. The experimental cosmic and perpendicular CCB data show high tracking efficiency and sub-millimeter residual widths. The tilted CCB configuration shows a significantly larger
\(\sigma^{\mathrm{LOO}}_z\), because the \(45^\circ\) track inclination in the \(yz\) projection spreads the ADC time profile over a larger timestamp interval, making the estimate of the drift coordinate more difficult. This motivated the sub-centroid study (see Section \ref{sec:pulse_shape_inclined_tracks}) to improve the extraction of drift co-ordinate $z$ that led to the drop of \(\sigma^{\mathrm{LOO}}_z\) of 0.549 mm and increase in the tracking efficiency to 95.52 $\%$.

Tables~\ref{tab:tpc_structure}--\ref{tab:tpc_charge_sharing} summarise the
structure of the reconstructed tracks and centroids. The number of centroids per event is proportional to the track multiplicity and to the amount of ionisation recorded in the event. Table~\ref{tab:tpc_adc_sample_usage} also shows the
fraction of centroid ADC that is used in fitted tracks, which indicates how much of the reconstructed signal is associated with the track fits. This percentage decreases due to high fluctuations in ionisation, background tracks or more complex track topologies produce centroids that are not assigned to fitted tracks. The cluster time extent shows how the drift conditions and track inclination affect the duration of the ADC time profile. The charge sharing between neighbouring pad columns is central to testing the zigzagreadout concept.

\begin{table}[H]
\centering
\small
\renewcommand{\arraystretch}{1.15}
\setlength{\tabcolsep}{4pt}
\caption{Structure of reconstructed tracks and centroids.}
\label{tab:tpc_structure}
\begin{tabularx}{\textwidth}{>{\raggedright\arraybackslash}p{3.5cm}CCC}
\hline
Data &
\shortstack{Centroids\\per event} &
\shortstack{Cluster width\\(pad columns)} &
\shortstack{Cluster time\\extent (timestamps)} \\
\hline
Simulation, perpendicular
& 10.28 & 3.05 & 9.45 \\
\hline
Cosmic 500 V/cm
& 12.47 & 3.05 & 8.11 \\

Cosmic 243 V/cm
& 11.41 & 3.91 & 9.88 \\

CCB perpendicular
& 19.51 & 2.58 & 7.21 \\

CCB tilted \(45^\circ\)
& 14.24 & 2.73 & 10.58 \\

CCB perpendicular higher-multiplicity
& 57.77 & 2.67 & 7.13 \\
\hline
\end{tabularx}
\end{table}

\begin{table}[H]
\centering
\small
\renewcommand{\arraystretch}{1.15}
\setlength{\tabcolsep}{4pt}
\caption{Summary of ADC-sample usage and clustering efficiency in the reconstructed centroids.}
\label{tab:tpc_adc_sample_usage}
\begin{tabularx}{\textwidth}{>{\raggedright\arraybackslash}p{3.8cm}CCC}
\hline
Data &
\shortstack{ADC in fitted\\centroids / all\\centroid ADC [\%]} &
\shortstack{Pads per\\active row} &
\shortstack{ADC samples assigned\\to centroids [\%]} \\
\hline
Simulation, perpendicular & 98.82 & 3.12 & 41.72 \\
\hline
Cosmic 500 V/cm & 95.82 & 3.74 & 31.63 \\
Cosmic 243 V/cm & 95.70 & 3.44 & 42.90 \\
CCB perpendicular & 84.32 & 4.77 & 63.06 \\
CCB tilted \(45^\circ\) & 56.00 & 4.11 & 67.35 \\
CCB perpendicular higher-multiplicity & 88.87 & 12.26 & 51.48 \\
\hline
\end{tabularx}
\end{table}

\begin{table}[H]
\centering
\small
\renewcommand{\arraystretch}{1.15}
\setlength{\tabcolsep}{5pt}
\caption{Summary of charge sharing in the reconstructed centroids.}
\label{tab:tpc_charge_sharing}
\begin{tabularx}{0.82\textwidth}{>{\raggedright\arraybackslash}p{3.8cm}CC}
\hline
Data &
\shortstack{Centroids with \(\geq 3\)\\pad columns [\%]} &
\shortstack{Mean\\\(N_{\mathrm{eff}}\)} \\
\hline
Simulation, perpendicular & 90.51 & 2.02 \\
\hline
Cosmic 500 V/cm & 79.96 & 2.27 \\
Cosmic 243 V/cm & 94.02 & 2.21 \\
CCB perpendicular & 53.96 & 2.06 \\
CCB tilted \(45^\circ\) & 65.13 & 2.01 \\
CCB perpendicular higher-multiplicity & 58.38 & 2.09 \\
\hline
\end{tabularx}
\end{table}

The effective number of pad columns, \(N_{\mathrm{eff}}\), was used to quantify the degree of charge sharing in the centroid calculation. It was calculated from the ADC amplitudes \(q_i\) in the pad columns contributing to a centroid as
\[
N_{\mathrm{eff}} = \frac{\left(\sum_i q_i\right)^2}{\sum_i q_i^2}.
\]
This quantity is equal to one when the centroid is dominated by a single pad column, and increases when the charge is shared more evenly between neighbouring pad columns. It accounts for how the charge is distributed among those columns.

\begin{figure}[H]
    \centering
    \begin{subfigure}{0.48\textwidth}
        \centering
        \includegraphics[width=\textwidth]{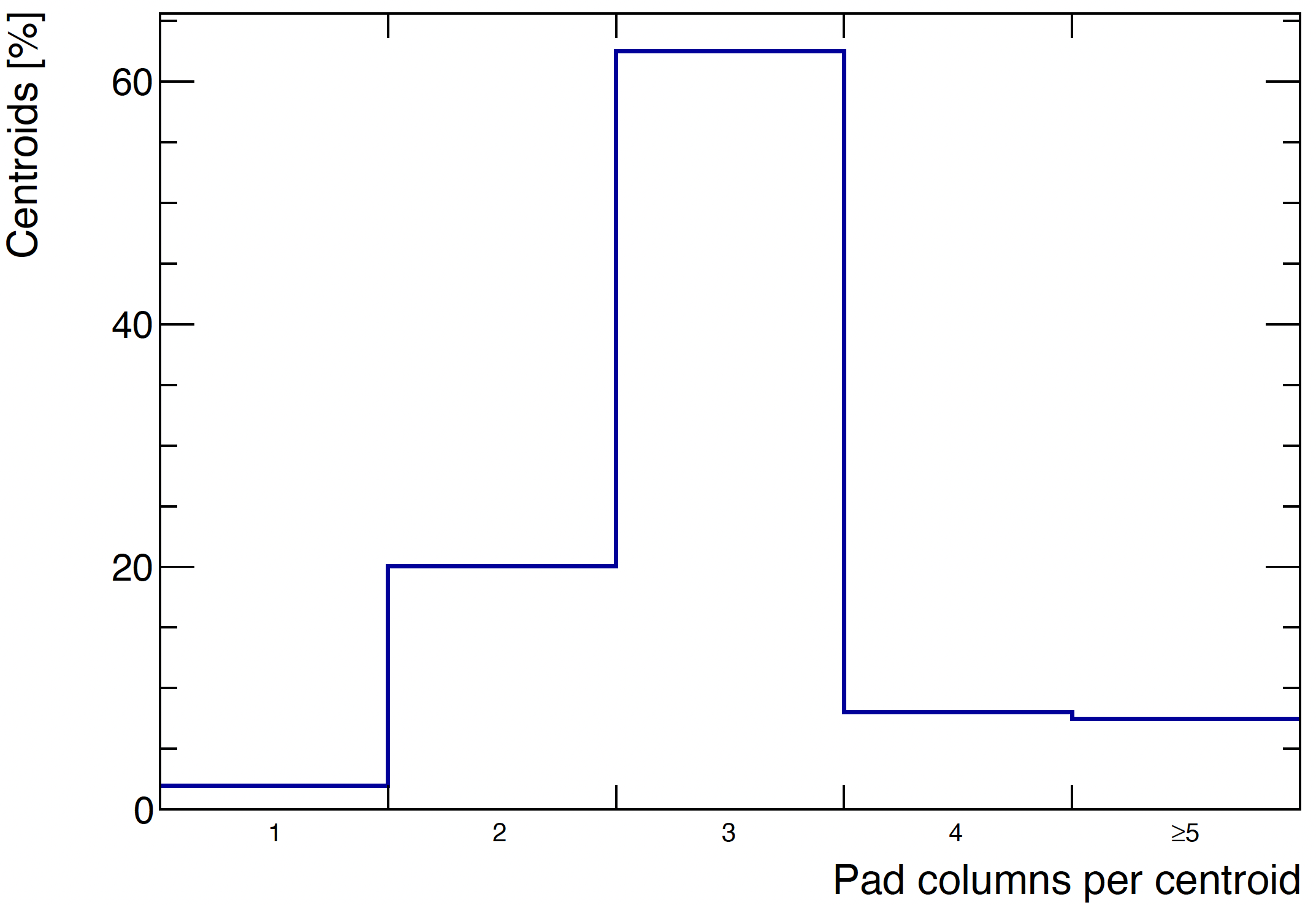}
        \caption{Cosmic measurement.}
        \label{fig:pad_columns_cosmic_500Vcm}
    \end{subfigure}
    \hfill
    \begin{subfigure}{0.48\textwidth}
        \centering
        \includegraphics[width=\textwidth]{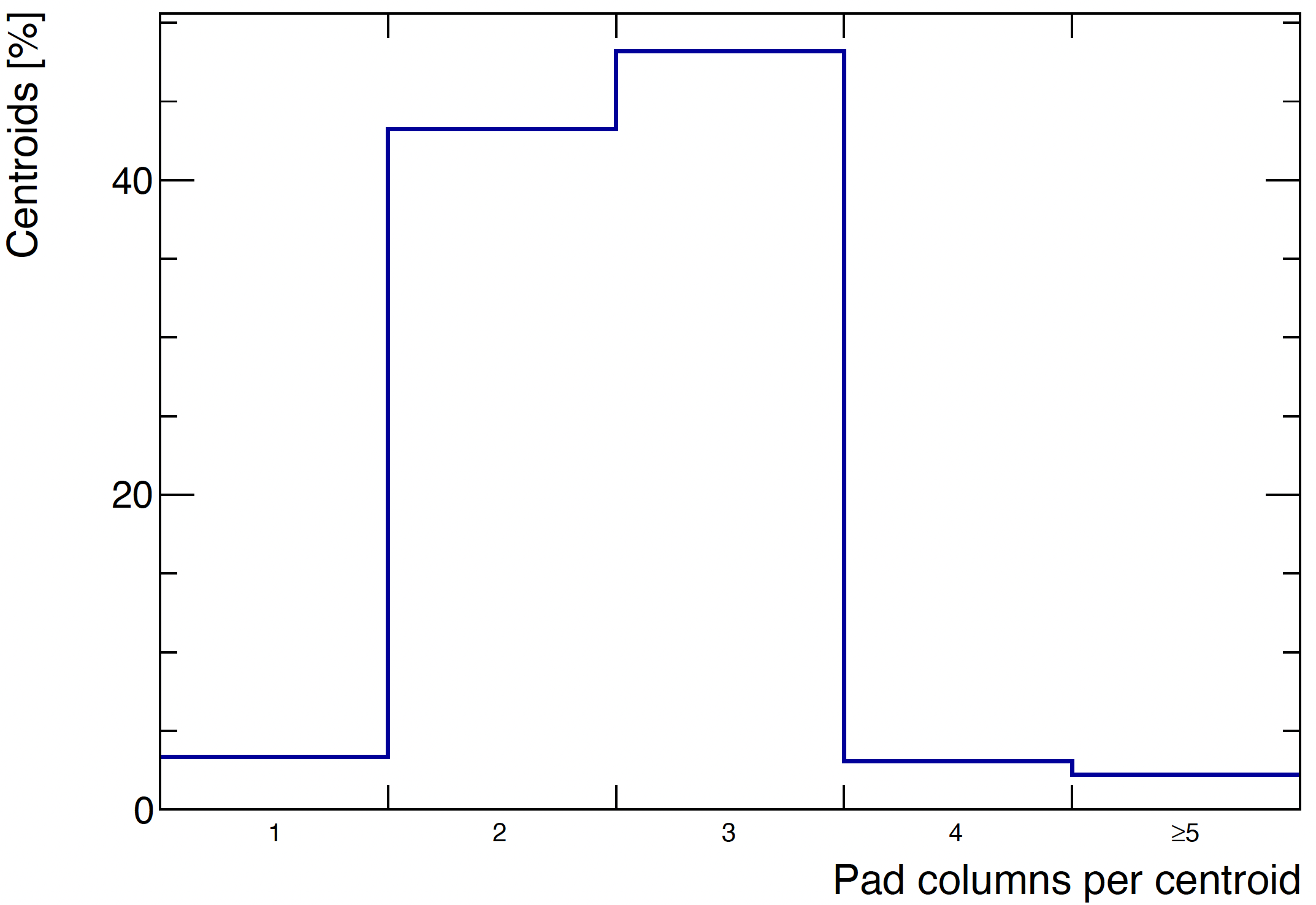}
        \caption{CCB perpendicular configuration.}
        \label{fig:pad_columns_ccb_perpendicular}
    \end{subfigure}
    \caption{Pad-column multiplicity of reconstructed centroids.}
    \label{fig:pad_columns_comparison}
\end{figure}

Figure~\ref{fig:pad_columns_comparison} compares the number of pad columns contributing to each reconstructed centroid for the cosmic measurement and for the CCB perpendicular configuration. In the cosmic data at \(E_{\mathrm{drift}}=500~\mathrm{V/cm}\), the distribution is dominated by centroids formed from three pad columns. In the CCB perpendicular configuration, the distribution shifts towards a larger fraction of two-pad centroids. This is consistent with the different track population in the CCB baseline measurement: the tracks are mostly close to perpendicular, with a smaller spread in the
transverse direction. 

\section{Detector Response in Cosmic Measurements}
\label{sec:cosmic_detector_response}

The cosmic measurements were used to characterise the response of the TPC prototype under several operating conditions and to validate the track-reconstruction procedure with experimental data. The study included variations of the GEM voltage, drift field and shaping time, and examined their effect on the energy deposited in terms of the \(dE/dx\) distribution, residual distributions, the number of reconstructed centroids per track and the charge sharing between neighbouring zigzagpads. These measurements therefore provide both a detector-response study and a reference sample for evaluating the tracking performance. Beam data taken at CCB were recorded under different trigger, rate and track-topology conditions and are treated separately in Section~\ref{sec:ccb_reconstruction_studies}.

The prototype response and its dependence on operating parameters were studied in detail in Ref.~\cite{Hehl2025_BSc}. That study examined the effect of the scintillator-trigger configuration, GEM voltage, shaping time and drift field. Here, the aspects most relevant for the track-reconstruction studies are revisited. The use of three scintillators was found to reduce the fraction of empty triggers compared with a two-scintillator trigger. The same study showed that \(V_{\mathrm{GEM}}=350~\mathrm{V}\) per GEM provided a suitable compromise between charge sharing and ADC saturation, while a shaping time of \(60~\mathrm{ns}\) still gave satisfactory \(dE/dx\) and residual distributions. Reducing the shaping time from \(120\) to \(60~\mathrm{ns}\) lowered the reconstructed \(dE/dx\) scale by approximately a factor of two, as expected from the shorter shaped pulses. The effect of the drift field and shaping time on the ADC time profile is discussed in Section~\ref{sec:cosmic_adc_time_profile_response}.

\begin{figure}[H]
  \centering
  \begin{subfigure}{.49\textwidth}
    \centering
    \includegraphics[width=\linewidth]{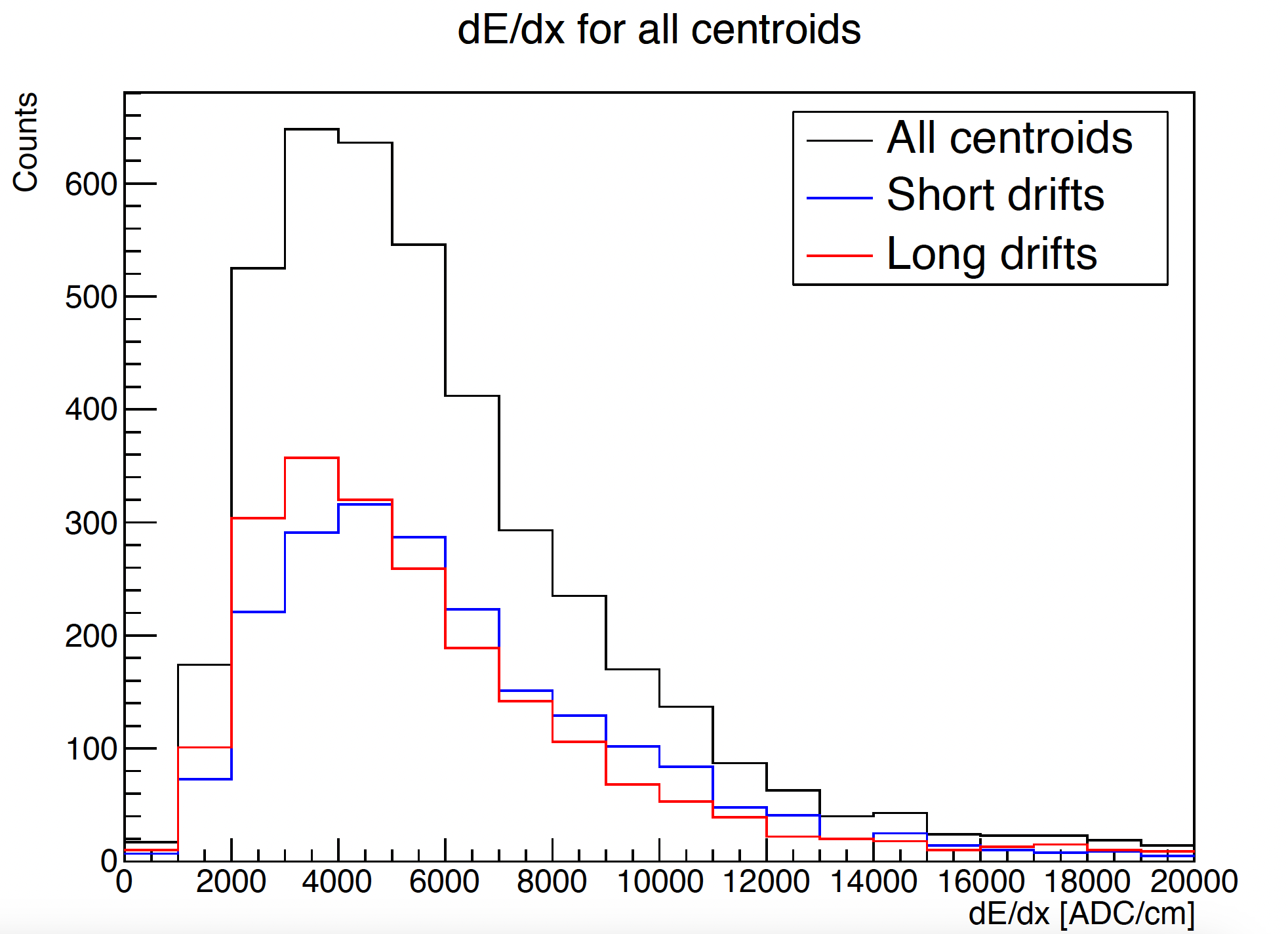}
    \caption{\(V_{\mathrm{GEM}}=350~\mathrm{V}\).}
    \label{fig:dEdX_CCB_3Dec_350V_GEM}
  \end{subfigure}
  \hfill
  \begin{subfigure}{.49\textwidth}
    \centering
    \includegraphics[width=\linewidth]{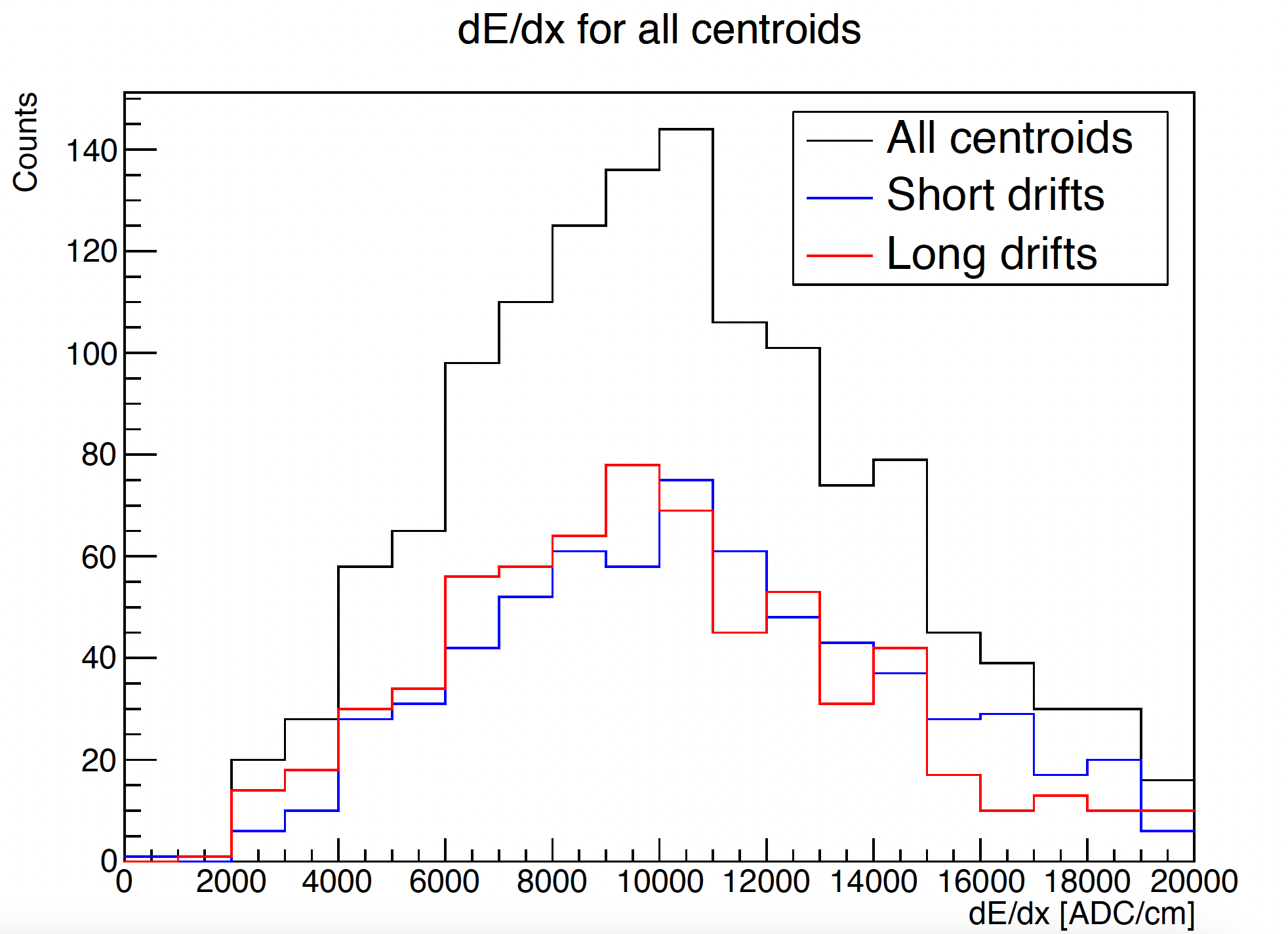}
    \caption{\(V_{\mathrm{GEM}}=360~\mathrm{V}\).}
    \label{fig:dEdX_CCB_3Dec_360V_GEM}
  \end{subfigure}
  \caption{Example \(dE/dx\) distributions for two GEM-voltage settings in
  cosmic data. The \(350~\mathrm{V}\) setting gives a clearer
  Landau-like peak, while the higher gain at \(360~\mathrm{V}\) broadens the
  distribution.}
  \label{fig:dEdX_GEM_Voltages}
\end{figure}

To check the GEM-voltage dependence with the approximate energy deposited,
size of the charge cloud and other tracking metrics, four settings were tested,
\[
V_{\mathrm{GEM}}=\{330,340,350,360\}~\mathrm{V}
\]
per GEM. For each voltage, the \(dE/dx\) distribution, the residual widths in the pad and drift directions, the average number of centroids per track, the
average number of fired pads contributing to each centroid and the ADC saturation fraction were determined. Here, the saturation fraction
\(f_{\mathrm{sat}}^{\mathrm{fit}}\) is defined as the fraction of ADC samples belonging to centroids used in fitted tracks with an ADC value above 900.
Example \(dE/dx\) spectra for \(V_{\mathrm{GEM}}=350~\mathrm{V}\) and
\(360~\mathrm{V}\) are shown in Figure~\ref{fig:dEdX_GEM_Voltages}, and the
corresponding summary values are listed in Table~\ref{tab:gem-scan}.

The mean \(\langle dE/dx\rangle\) decreases by approximately a factor of two
for each \(10~\mathrm{V}\) reduction of \(V_{\mathrm{GEM}}\), consistent with
the strong dependence of the gas gain on the GEM voltage. The residual widths
remain below \(0.5~\mathrm{mm}\) for all tested voltages, and the average
number of reconstructed centroids remains close to ten. This indicates that
tracks can be reconstructed across the full scan. The main practical difference
between the settings is observed in the charge sharing, the shape of the
\(dE/dx\) distribution and the onset of ADC saturation. At \(360~\mathrm{V}\),
the average number of fired pads per centroid increases to
\(\langle N_{\mathrm{pads}}\rangle=3.17\), and the \(dE/dx\) distribution
becomes broader. The fraction of ADC samples above the saturation threshold
also increases to \(f_{\mathrm{sat}}^{\mathrm{fit}}=19.54\%\). At
\(350~\mathrm{V}\), the charge sharing remains close to the desired range of two to three pads per centroid, \(\langle N_{\mathrm{pads}}\rangle=2.88\),
while the \(dE/dx\) distribution retains a clearer Landau-like peak and the saturation fraction is reduced to \(f_{\mathrm{sat}}^{\mathrm{fit}}=6.09\%\). The setting \(V_{\mathrm{GEM}}=350~\mathrm{V}\) was therefore used as the
reference GEM voltage for the cosmic reconstruction studies.

\begin{table}[tb]
  \centering
  \caption{Summary of the GEM-voltage scan for cosmic data. The table
  lists the mean \(dE/dx\), the residual widths in the pad (\(x\)) and drift
  (\(z\)) directions, the average number of reconstructed centroids per track,
  the average number of fired pads contributing to each centroid and the ADC
  saturation fraction \(f_{\mathrm{sat}}^{\mathrm{fit}}\). A sample was counted
  as saturated if its ADC value was above 900.}
  \label{tab:gem-scan}
  \begin{tabular}{c c c c c c c}
    \hline
    \(V_{\mathrm{GEM}}\) [V] &
    \(\langle dE/dx\rangle\) [ADC/cm] &
    \(\sigma_x\) [mm] &
    \(\sigma_z\) [mm] &
    \(\langle N_{\mathrm{cent}}\rangle\) &
    \(\langle N_{\mathrm{pads}}\rangle\) &
    \(f_{\mathrm{sat}}^{\mathrm{fit}}\) [\%] \\
    \hline
    360 & 10290 & 0.55 & 0.42 & 9.97 & 3.17 & 19.54 \\
    350 &  5909 & 0.51 & 0.28 & 9.94 & 2.88 &  6.09 \\
    340 &  3036 & 0.52 & 0.27 & 9.97 & 2.76 &  1.55 \\
    330 &  1594 & 0.55 & 0.27 & 9.92 & 2.70 &  0.33 \\
    \hline
  \end{tabular}
\end{table}

\begin{figure}[H]
  \centering
  \includegraphics[width=0.60\textwidth]{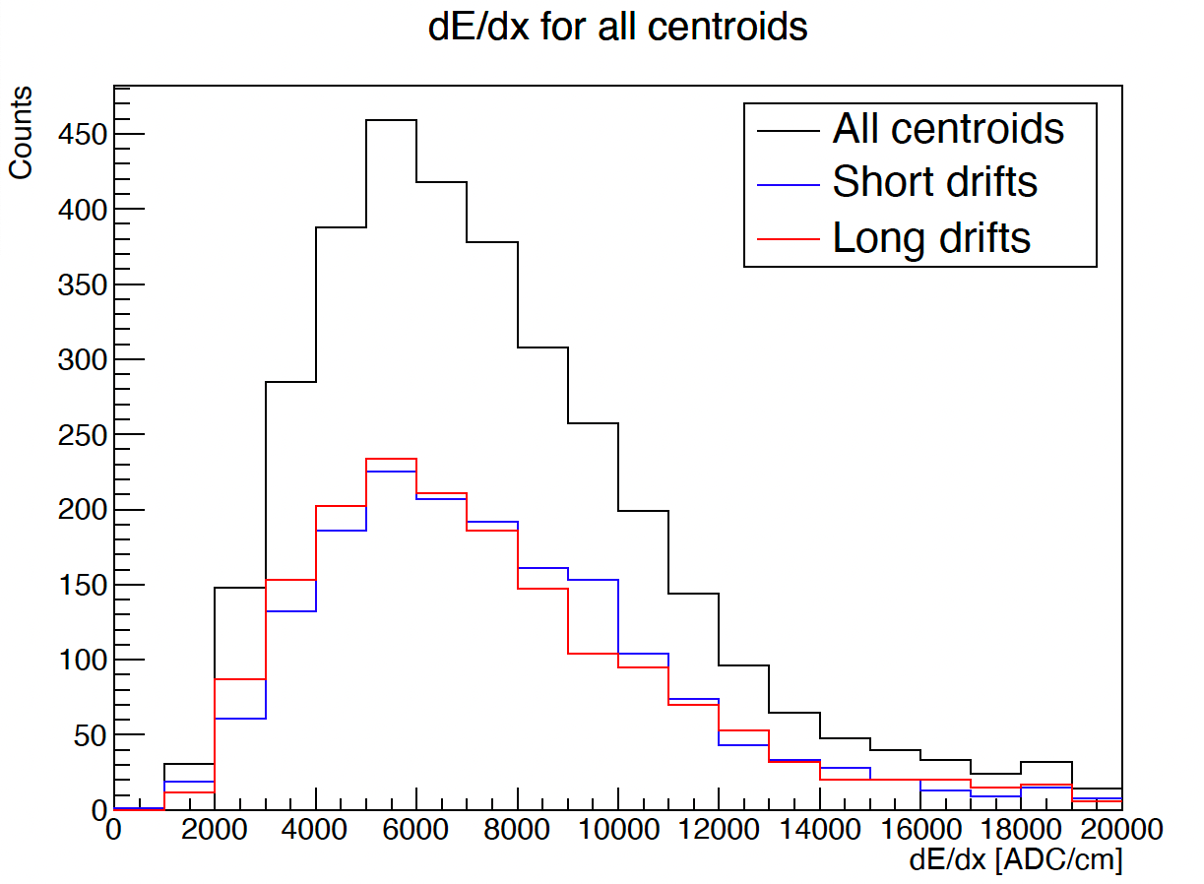}
  \caption{Comparison of the reconstructed \(dE/dx\) distributions for short and long drift distances at the reference cosmic operating point, \(E_{\mathrm{drift}}=500~\mathrm{V/cm}\) and gas flow \(12~\mathrm{l/h}\). The two distributions nearly overlap, indicating stable charge collection across the drift volume used in the analysis.}
  \label{fig:500Vcm_high_gasflow}
\end{figure}

The uniformity of the collected charge was also checked for the reference cosmic operating conditions. For this purpose, reconstructed centroids were split into short- and long-drift samples using the median reconstructed drift length, so that both samples contained the same number of entries. For each centroid, \(dE/dx\) was calculated as the collected charge divided by the local track length. At \(E_{\mathrm{drift}}=500~\mathrm{V/cm}\) and a gas flow of \(12~\mathrm{l/h}\), the short- and long-drift \(dE/dx\) distributions nearly overlap, as shown in Figure~\ref{fig:500Vcm_high_gasflow}. The corresponding mean values are \(7568.08~\mathrm{ADC/cm}\) and \(7412.66~\mathrm{ADC/cm}\), respectively. This indicates that no strong drift-length-dependent charge loss is observed under the operating conditions used for the cosmic reconstruction study. 

\begin{figure}[H]
  \centering
  \begin{subfigure}{.49\textwidth}
    \centering
    \includegraphics[width=\linewidth]{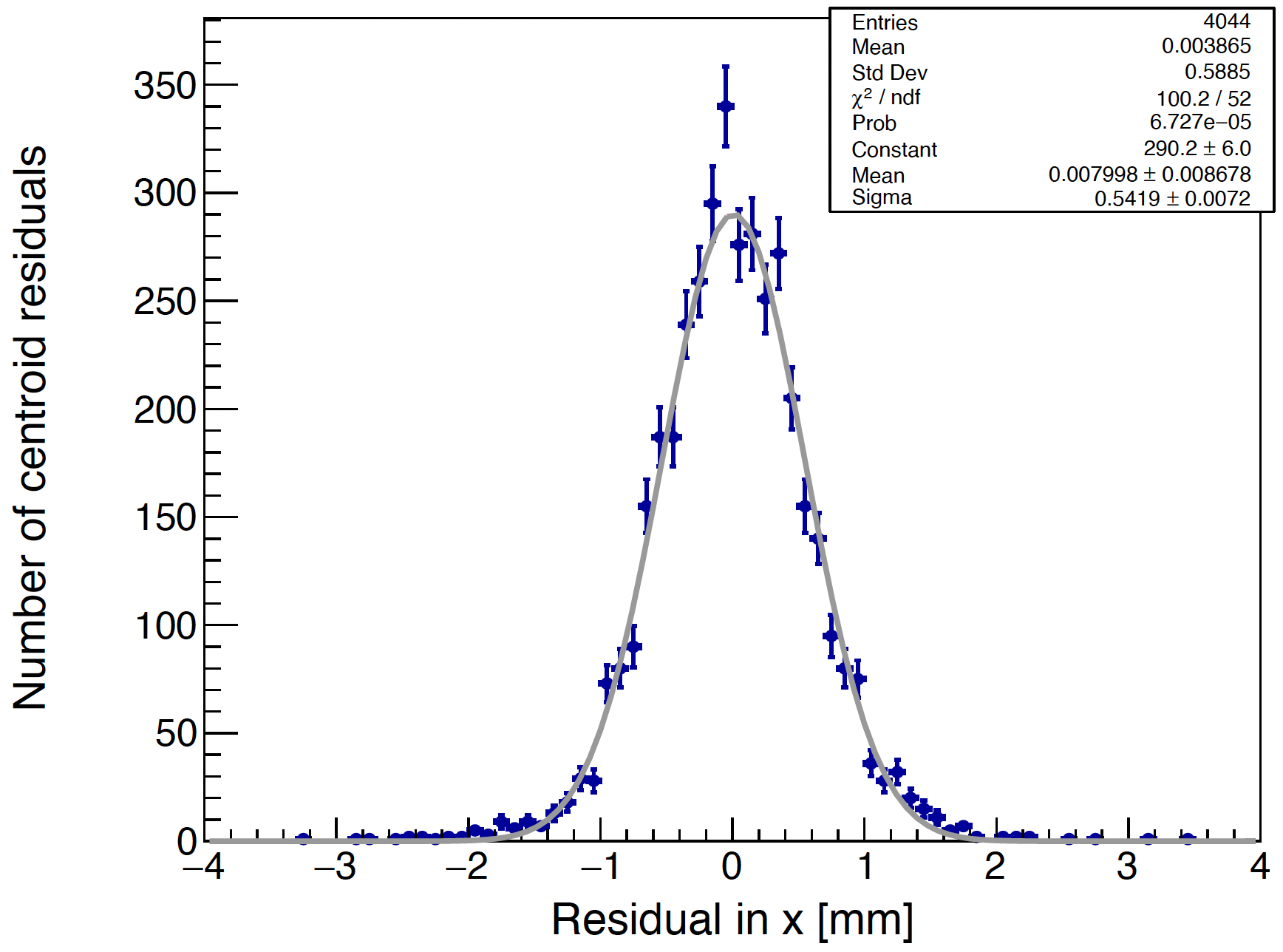}
    \caption{\(yx\) projection.}
    \label{fig:20250925_x_residuals}
  \end{subfigure}
  \hfill
  \begin{subfigure}{.49\textwidth}
    \centering
    \includegraphics[width=\linewidth]{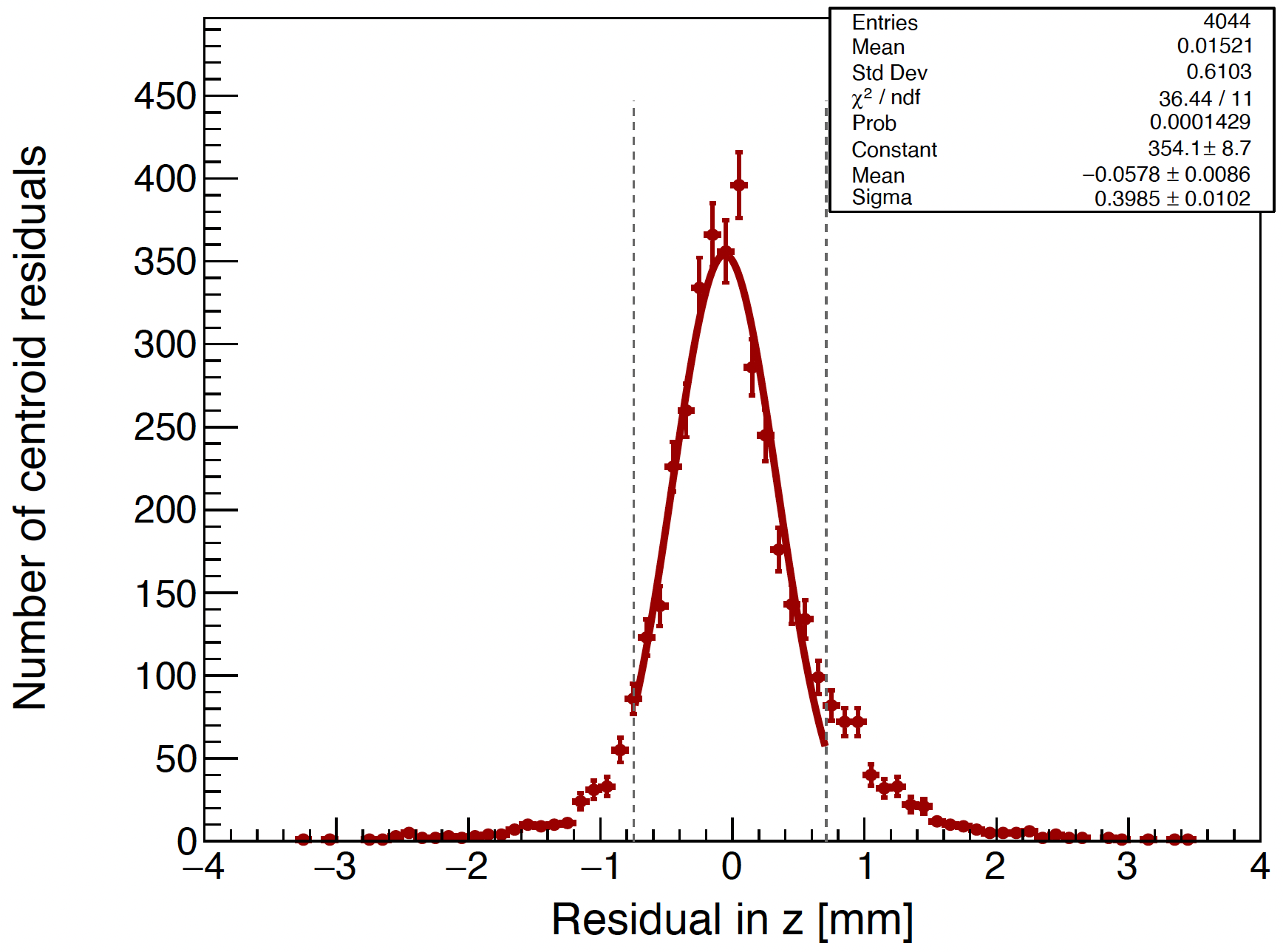}
    \caption{\(yz\) projection.}
    \label{fig:20250925_z_residuals}
  \end{subfigure}
  \caption{Representative residual distributions for tracks reconstructed at the reference cosmic operating point. The distributions are centred close to zero and have sub-millimetre fitted widths. In the \(yz\) projection, the Gaussian fit is restricted to the central core because of the asymmetric tails in the residual distribution.}
  \label{fig:20250925_residuals}
\end{figure}

A detailed comparison of residual widths for the different data sets is given in Section~\ref{sec:validation_application}. Here, one representative residual distribution is shown for the reference cosmic operating point as a visual check of the residual-distribution shape. The residual distributions in the \(yx\) and \(yz\) projections are shown in Figure~\ref{fig:20250925_residuals}. They are centred close to zero and have fitted widths of approximately \(\sigma_x=0.54~\mathrm{mm}\) and \(\sigma_z=0.40~\mathrm{mm}\), respectively. For the \(yz\) projection, the Gaussian fit was restricted to the central core of the distribution, since asymmetric tails are visible. These tails are consistent with a less precise extraction of the time coordinate for inclined tracks present in the data set. The fitted widths indicate sub-millimetre residuals for the selected cosmic configuration.

\begin{figure}[H]
  \centering
  \includegraphics[width=0.62\textwidth]{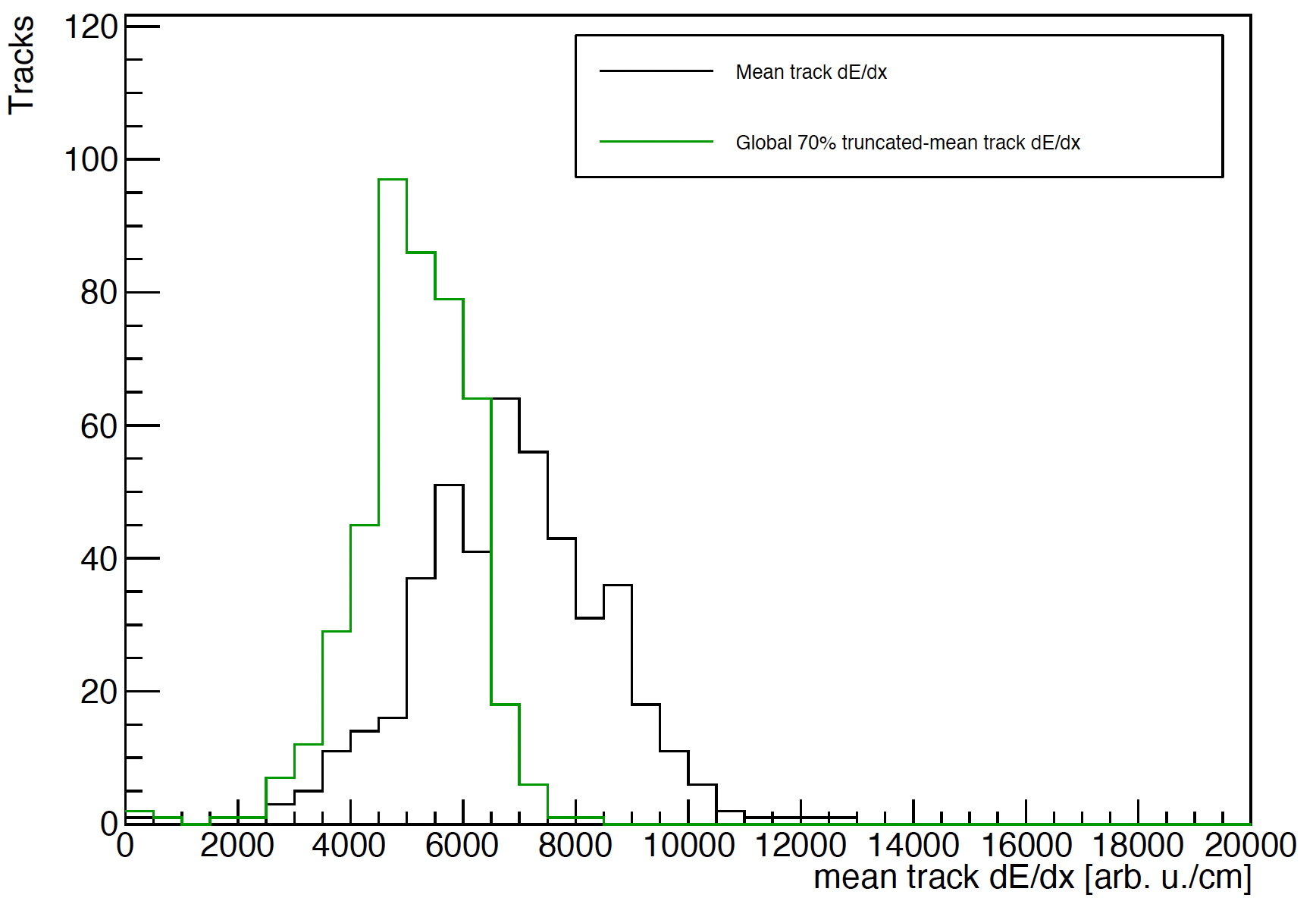}
  \caption{Track-level \(dE/dx\) distributions for the reference cosmic data set,
  comparing the arithmetic mean over all centroids with the truncated mean
  after applying the global 70th-percentile centroid-level threshold.}
  \label{fig:dedx_truncated_mean_cosmic}
\end{figure}

The same reference cosmic data set was also used to estimate a track-level
\(dE/dx\) resolution relevant for particle-identification studies. Since the
sample is expected to be dominated by minimum-ionising cosmic muons, it
provides a useful reference for the minimum-ionising response of the prototype.
For each reconstructed track, the centroid-level \(dE/dx\) values were combined
into a single track-level value. The arithmetic mean used all centroids assigned
to the track, while the truncated mean excluded centroid values above the global 70th-percentile threshold of the centroid-level \(dE/dx\) distribution. This reduces the influence of large local ionisation fluctuations, such as those
associated with \(\delta\)-electrons. As shown in
Figure~\ref{fig:dedx_truncated_mean_cosmic}, the truncated-mean distribution
is narrower and closer to a Gaussian shape than the arithmetic-mean
distribution. A Gaussian fit gives a relative width of approximately
\(\sigma/\mu \simeq 0.16\), interpreted here as the relative resolution of the
ADC-based track-level \(dE/dx\) value for the minimum-ionising reference
dataset.

\subsection{ADC time-profile response}
\label{sec:cosmic_adc_time_profile_response}

\begin{figure}[H]
  \centering
  \begin{subfigure}[t]{0.49\textwidth}
    \centering
    \includegraphics[width=\linewidth]{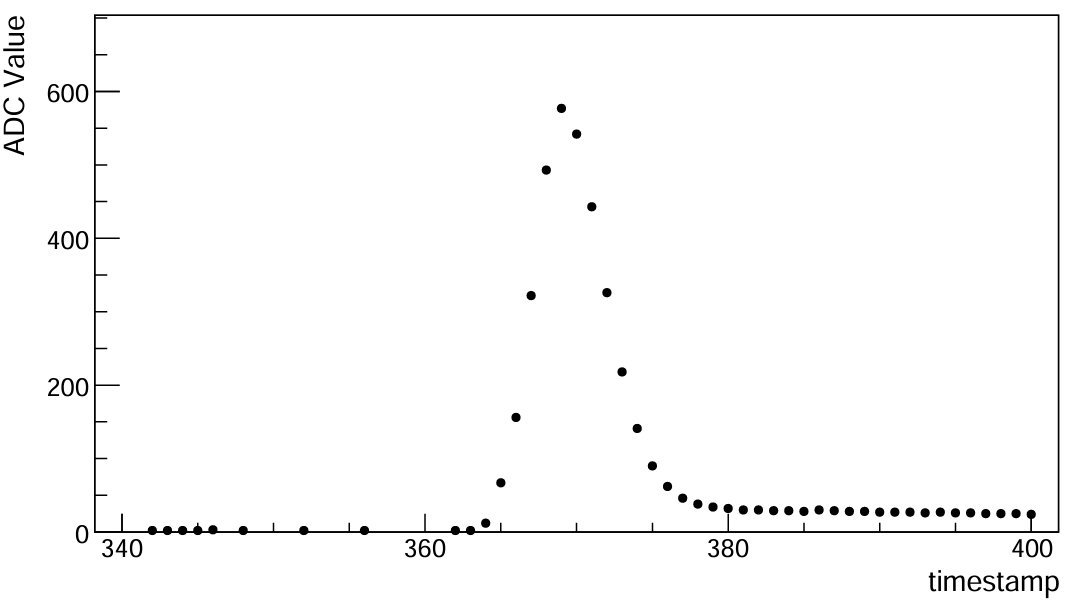}
    \caption{Central-pad-column time profile for \(E_{\mathrm{drift}}=243~\mathrm{V/cm}\), corresponding to \(v_{\mathrm{drift}}\simeq0.73~\mathrm{cm}/\mu\mathrm{s}\).}
    \label{fig:pulse_shape_243Vcm}
  \end{subfigure}
  \hfill
  \begin{subfigure}[t]{0.49\textwidth}
    \centering
    \includegraphics[width=\linewidth]{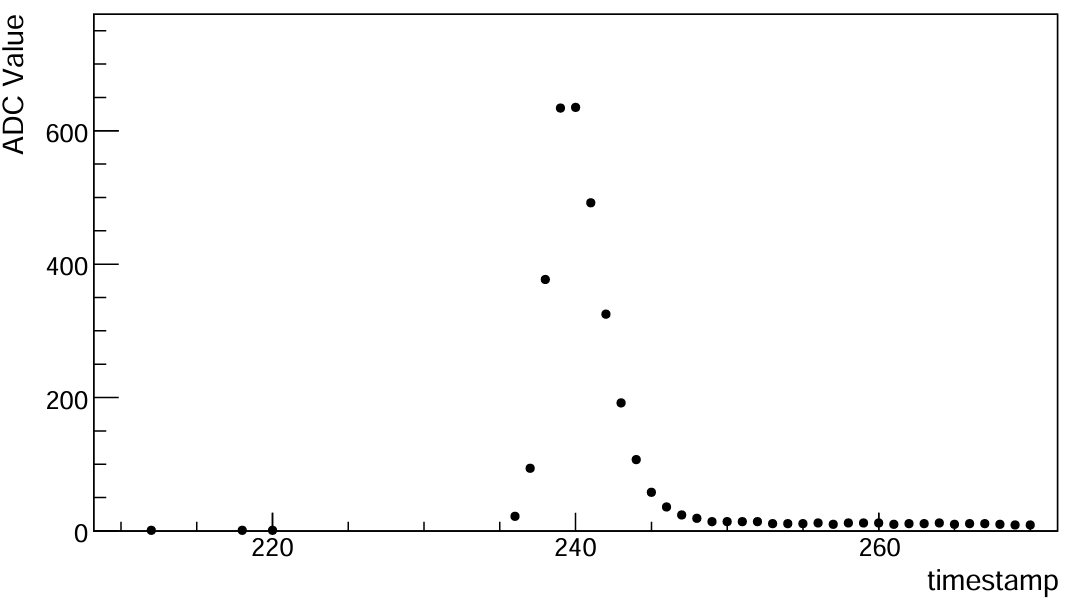}
    \caption{Central-pad-column time profile for \(E_{\mathrm{drift}}=500~\mathrm{V/cm}\), corresponding to \(v_{\mathrm{drift}}\simeq1.6~\mathrm{cm}/\mu\mathrm{s}\).}
    \label{fig:pulse_shape_500Vcm}
  \end{subfigure}

  \vspace{0.8em}

  \begin{subfigure}[t]{0.49\textwidth}
    \centering
    \includegraphics[width=\linewidth]{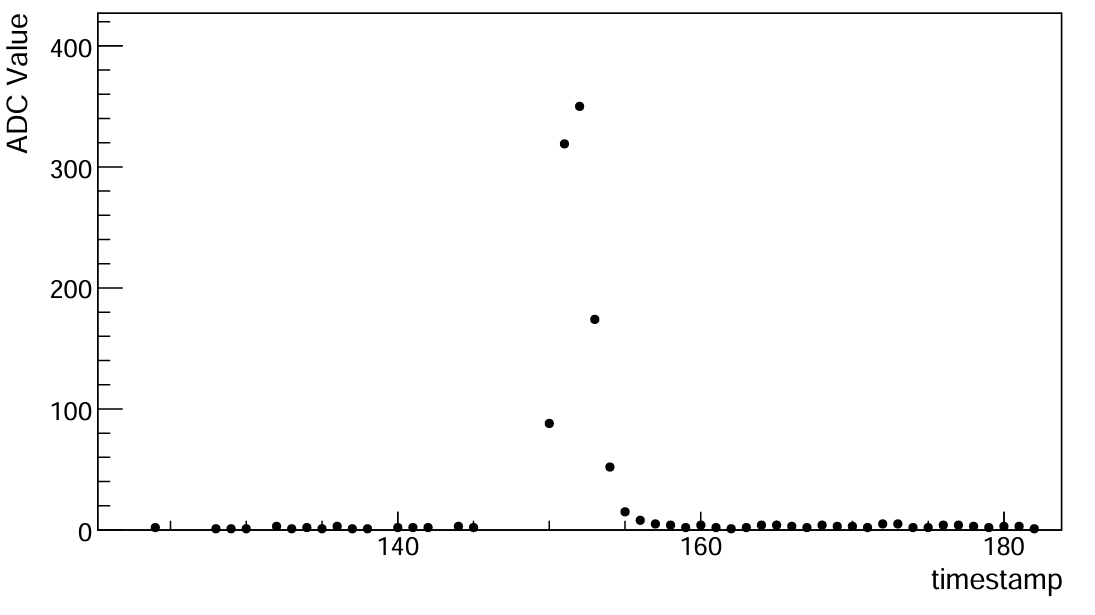}
    \caption{Central-pad-column time profile for a shaping time of \(60~\mathrm{ns}\).}
    \label{fig:pulse_shape_60ns}
  \end{subfigure}
  \hfill
  \begin{subfigure}[t]{0.49\textwidth}
    \centering
    \includegraphics[width=\linewidth]{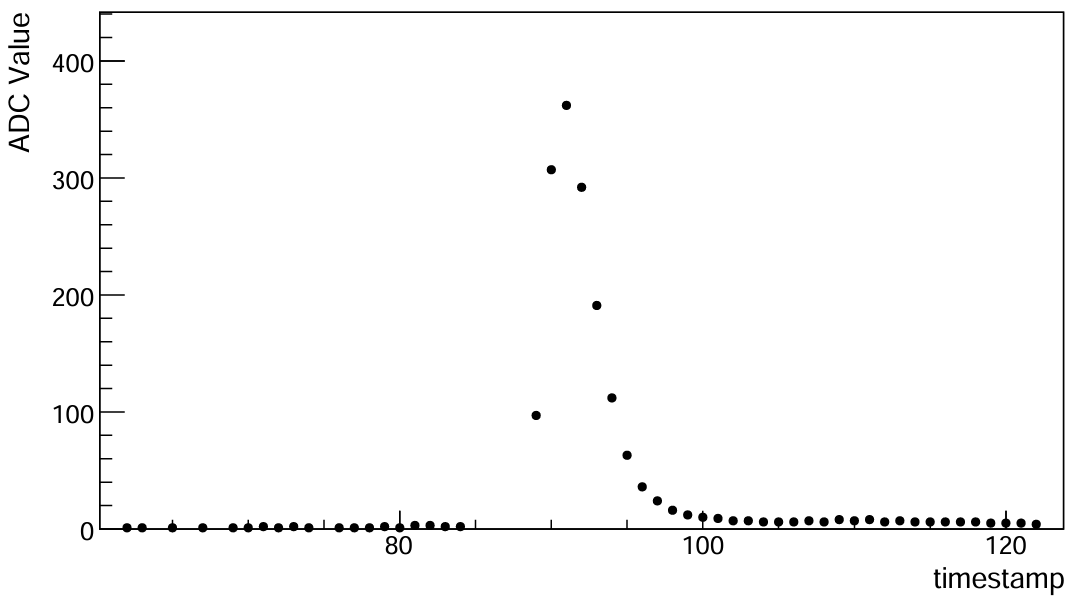}
    \caption{Central-pad-column time profile for a shaping time of \(120~\mathrm{ns}\).}
    \label{fig:pulse_shape_120ns}
  \end{subfigure}

  \caption{Representative central-pad-column time profiles from reconstructed clusters in cosmic measurements, illustrating the dependence of the ADC time profile on drift speed and shaping time. The lower drift speed increases the spread in timestamp, while the shaping time changes the readout response to the collected charge.}
  \label{fig:pulse_shape_drift_shaping_examples}
\end{figure}

The dependence of the observed pulse shape on detector settings was studied using cosmic measurements. Examples are shown in Figure~\ref{fig:pulse_shape_drift_shaping_examples}, where the ADC values are shown for the central pad column of a reconstructed cluster, defined as the pad column within that cluster with the largest integrated ADC value. This choice avoids mixing neighbouring pad columns and gives a compact view of the time profile associated with the largest charge contribution in the cluster. At the lower drift-field setting, the lower drift speed, \(v_{\mathrm{drift}}\), means that the same geometrical spread in drift distance corresponds to a broader spread in timestamp. The shaping time also changes the measured pulse profile: a shorter shaping time preserves more of the time structure of the arriving charge, whereas a longer shaping time produces a more smoother readout response.

\FloatBarrier
\newpage

\section{Reconstruction Studies with the CCB Proton Data}
\label{sec:ccb_reconstruction_studies}

\subsection{Energy-loss distributions in the CCB proton data}
\label{sec:ccb_energy_loss_distributions}

The CCB elastic-scattering data extend the validation beyond the cosmic measurements because the trigger-track sample is expected to contain protons with substantially larger ionisation than the particles dominating the cosmic data. As discussed in Section~\ref{sec:ccb_proton_test}, protons with kinetic energies of about \(30\)--\(40~\mathrm{MeV}\) at the TPC position are expected to release substantially more ionisation charge per unit length than minimum-ionising cosmic particles.

\begin{figure}[H]
    \centering
    \includegraphics[width=0.72\textwidth]{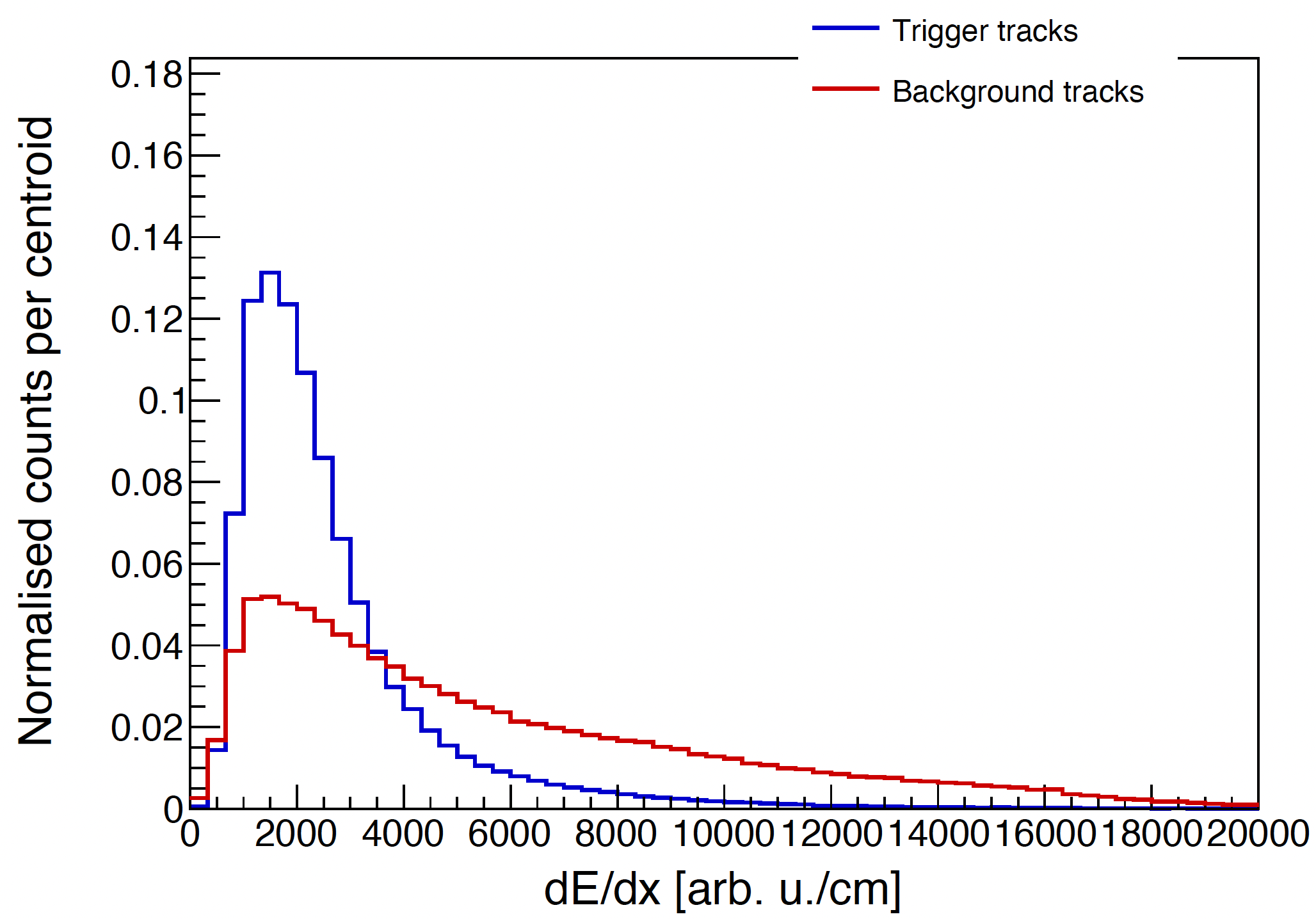}
    \caption{Centroid-normalised \(dE/dx\) distributions for trigger tracks and background tracks in the analysed CCB data set. Trigger tracks are defined as tracks whose projection to the TPC face falls inside the geometrical window defined by the trigger scintillators. Background tracks are reconstructed outside this window. The reconstructed \(dE/dx\) is used as a quantity proportional to the charge deposited per unit track length.}
    \label{fig:dedx_trigger_background_ccb}
\end{figure}

The reconstructed \(dE/dx\) is used here as a quantity proportional to the charge deposited per unit track length. Figure~\ref{fig:dedx_trigger_background_ccb} compares centroid-normalised \(dE/dx\) distributions for tracks whose projection to the TPC face falls inside the geometrical window defined by the trigger scintillators and for tracks reconstructed outside this window. The first group is referred to as trigger tracks, while the second group is referred to as background tracks. The comparison shows that the scintillator-defined geometrical selection isolates a distinct subset of tracks, while the background-track group has a broader high-\(dE/dx\) tail.

\subsection{Pulse-shape information for tracks inclined in the drift plane}
\label{sec:pulse_shape_inclined_tracks}

The dependence of the ADC time profile on detector operating conditions was discussed in Section~\ref{sec:cosmic_adc_time_profile_response}. Although inclined tracks are also present in the cosmic measurements, the dedicated tilted CCB runs provide a high-statistics track population with a similar inclination in the drift plane throughout the measurement. This makes the data set particularly useful for studying how the track geometry affects the pulse shape. For tracks inclined in the drift plane, the ionisation associated with one pad row can be spread over a larger range in the drift direction \(z\) than for nearly perpendicular tracks. Consequently, electrons from the same pad row can have different drift distances to the readout plane and arrive over a longer time interval, which broadens the cluster time profile.

This effect can be estimated directly from the geometry. For a track crossing a pad row of height \(10~\mathrm{mm}\) at an angle of \(45^\circ\) in the \(yz\) plane, the extent of the ionisation segment along the drift direction is
\begin{equation}
\Delta z = 10~\mathrm{mm}\tan 45^\circ = 10~\mathrm{mm}.
\end{equation}
The corresponding path length inside the pad row increases to
\begin{equation}
L = \frac{10~\mathrm{mm}}{\cos45^\circ} \approx 14.1~\mathrm{mm}.
\end{equation}
For a drift speed of \(v_{\mathrm{drift}}=2~\mathrm{cm/\mu s}\), this gives a geometrical contribution to the time-profile broadening of
\begin{equation}
\Delta t = \frac{\Delta z}{v_{\mathrm{drift}}}
= \frac{10~\mathrm{mm}}{20~\mathrm{mm/\mu s}}
= 0.5~\mathrm{\mu s},
\end{equation}
which corresponds to approximately ten timestamps for a \(20~\mathrm{MHz}\) sampling rate.

\begin{figure}[H]
  \centering
  \begin{subfigure}[t]{0.49\textwidth}
    \centering
    \includegraphics[width=\linewidth]{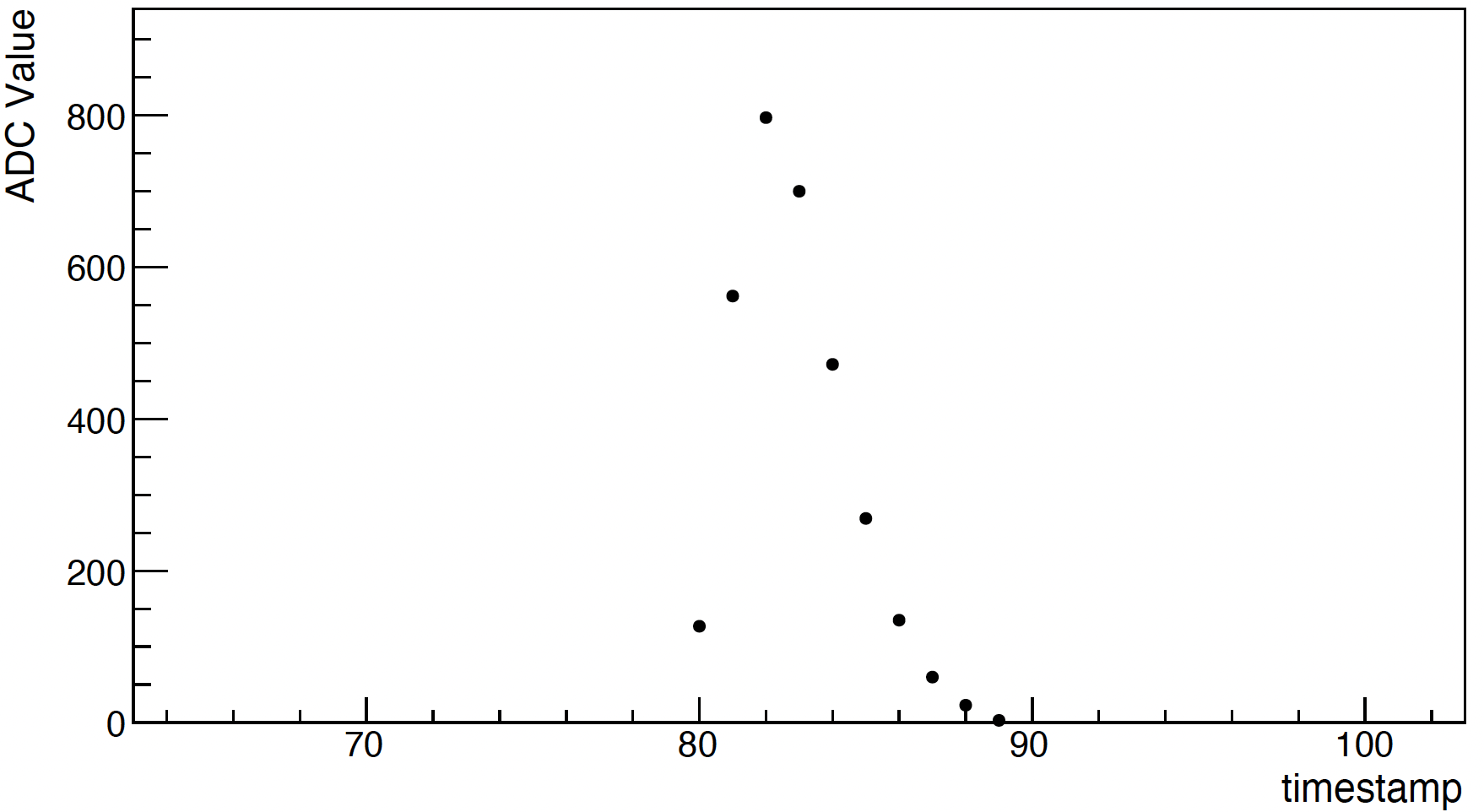}
    \caption{Central-pad-column time profile for the perpendicular-track CCB configuration.}
    \label{fig:pulse_shape_ccb_perpendicular}
  \end{subfigure}
  \hfill
  \begin{subfigure}[t]{0.49\textwidth}
    \centering
    \includegraphics[width=\linewidth]{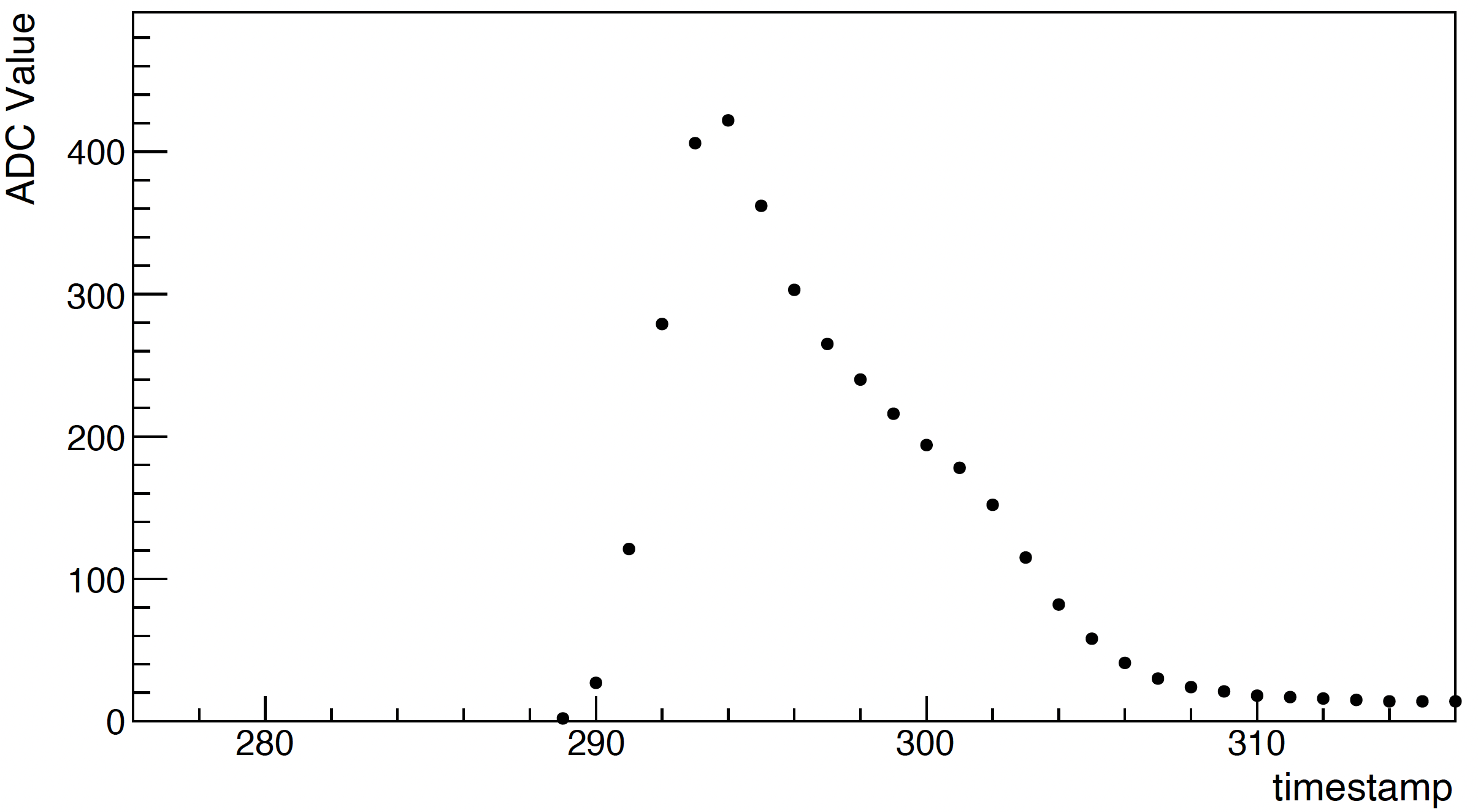}
    \caption{Central-pad-column time profile for the tilted-track CCB configuration.}
    \label{fig:pulse_shape_ccb_tilted}
  \end{subfigure}
  \caption{Representative central-pad-column time profiles from reconstructed clusters in the CCB data. In the tilted configuration, the track is inclined in the drift plane and the charge assigned to a single pad row is spread over a longer time interval than in the perpendicular-track configuration.}
  \label{fig:pulse_shape_ccb_perpendicular_tilted}
\end{figure}

The effect of the track geometry on the pulse shape is visible directly in the CCB data. Figure~\ref{fig:pulse_shape_ccb_perpendicular_tilted} compares representative central-pad-column time profiles from the perpendicular-track and tilted-track CCB configurations. In the perpendicular-track configuration, the charge in a pad row is concentrated in a narrow time interval. In the tilted configuration, where the TPC was rotated by \(45^\circ\) around the vertical \(x\) axis, the charge assigned to the same pad row can extend over a much longer time interval. This demonstrates the practical reconstruction problem: a single centroid timestamp may no longer represent the charge distribution in the pad row well.

The important point for reconstruction is that the time profile is not determined only by the electronics response. For tracks inclined in the drift plane, it also contains information about the longitudinal distribution of charge within a pad row. This motivates using the internal time structure of the pulse, rather than only a single charge-weighted timestamp, when reconstructing strongly inclined tracks.

\subsection{Sub-centroid reconstruction for tracks inclined in the drift plane}
\label{sec:subcentroid_inclined_tracks}

In the standard reconstruction, one centroid is assigned to each pad row and its \(y\) coordinate is placed at the centre of the pad row. This is reasonable for nearly perpendicular tracks, where the cluster time profile extends over a relatively small time range within one pad row. For inclined tracks, the situation is different. Different parts of the cluster time profile can
correspond to different local \(y\) positions within the same pad row.
Moreover, ionisation fluctuations can distort the time profile, making the
extraction of a representative timestamp more difficult. This effect mainly degrades the reconstruction in the \(yz\) projection, where the drift coordinate \(z\) is obtained from the time information.

\begin{figure}[H]
  \centering
  \begin{subfigure}[t]{0.49\textwidth}
    \centering
    \includegraphics[width=\linewidth]{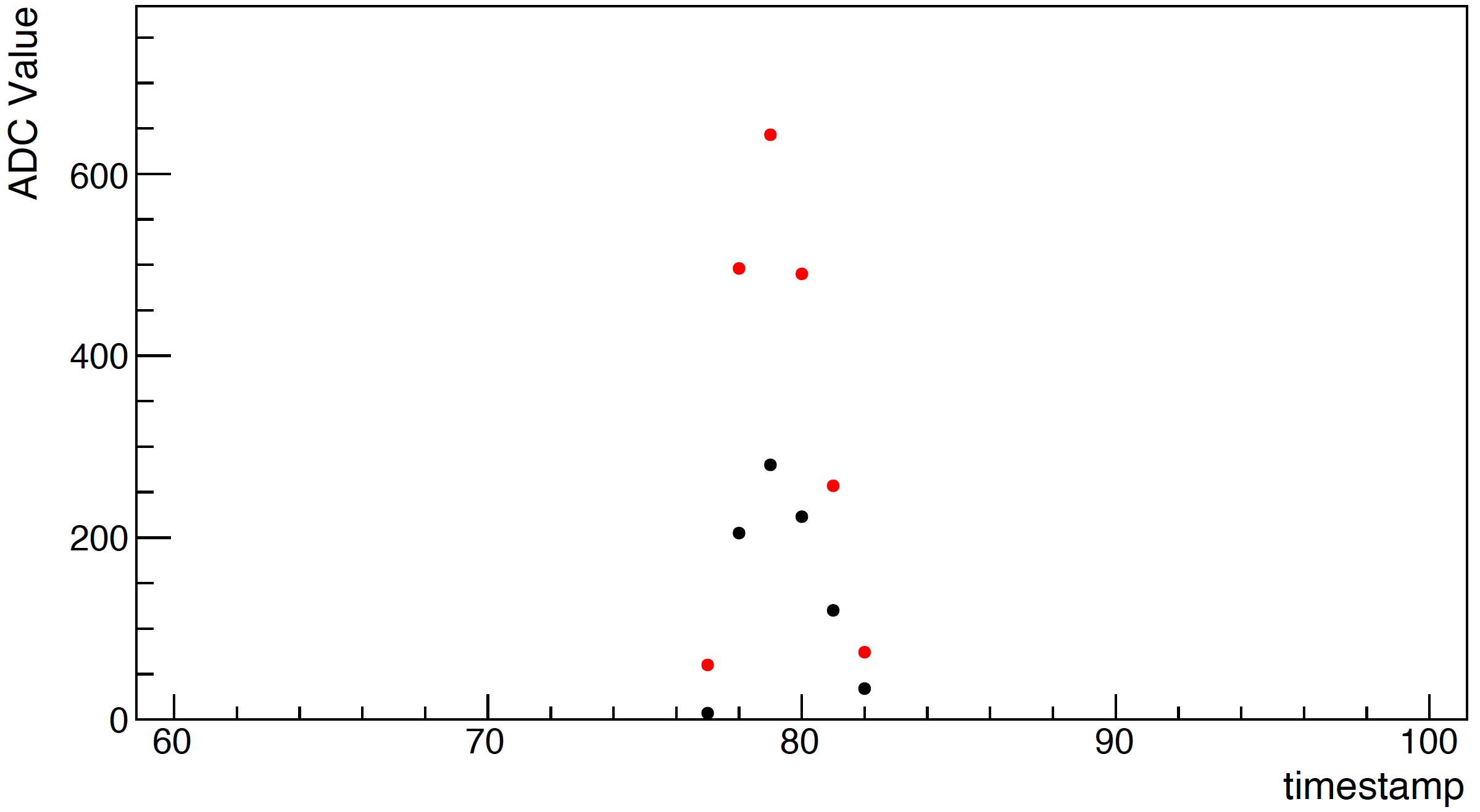}
    \caption{Perpendicular-track CCB configuration.}
    \label{fig:pulse_shape_ccb_adjacent_rows_perpendicular}
  \end{subfigure}
  \hfill
  \begin{subfigure}[t]{0.49\textwidth}
    \centering
    \includegraphics[width=\linewidth]{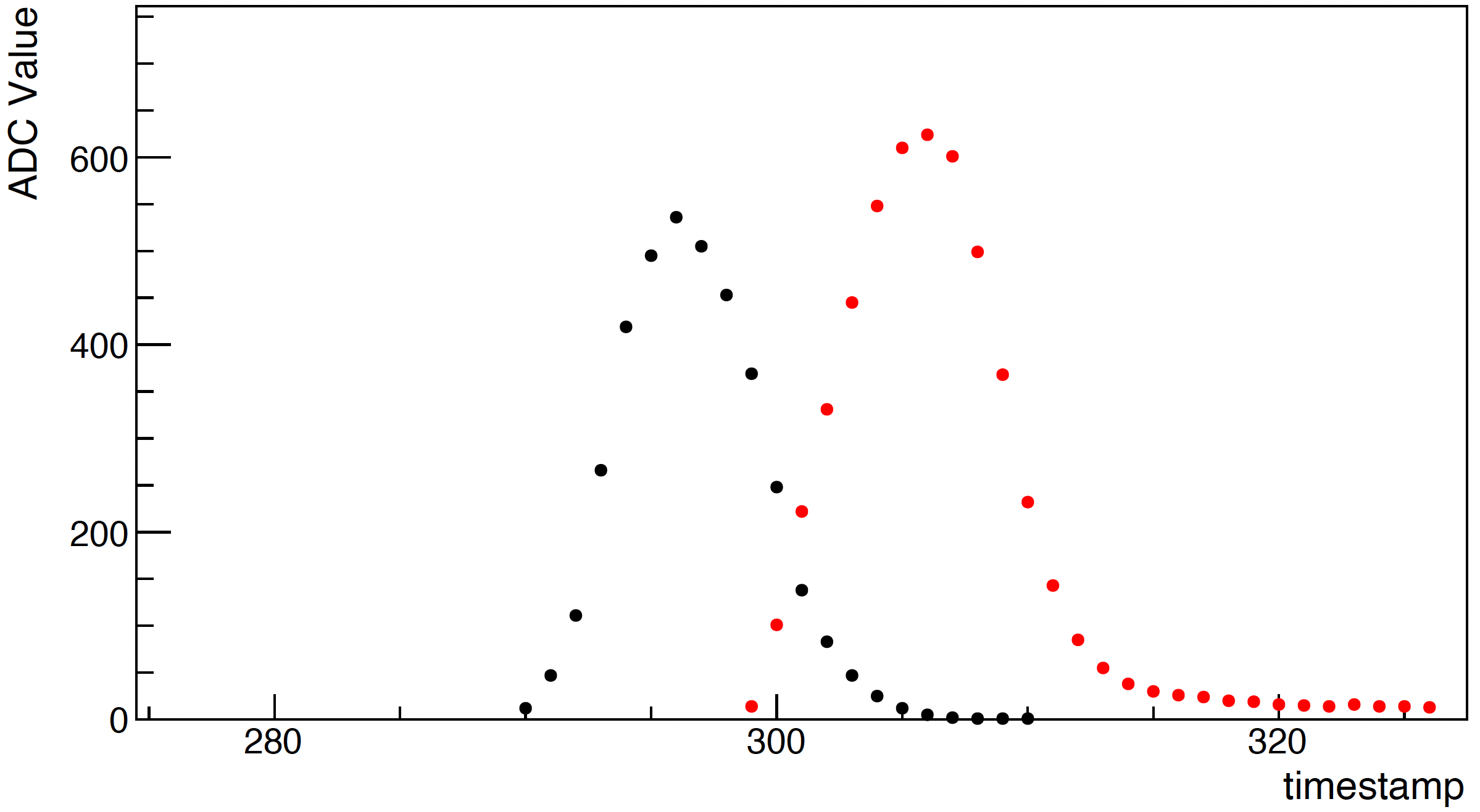}
    \caption{Tilted-track CCB configuration.}
    \label{fig:pulse_shape_ccb_adjacent_rows_tilted}
  \end{subfigure}
  \caption{Comparison of summed cluster time profiles in neighbouring pad rows for representative CCB events. The black points show the profile in row 1 and the red points show the profile in row 2. In the perpendicular-track configuration the profiles are nearly aligned, while in the tilted configuration the neighbouring row is shifted to later timestamps.}
  \label{fig:pulse_shape_ccb_adjacent_rows}
\end{figure}

To show this effect directly, the summed cluster time profiles were compared for two neighbouring pad rows. Figure~\ref{fig:pulse_shape_ccb_adjacent_rows} shows this comparison for representative perpendicular-track and tilted-track CCB events. In the perpendicular-track configuration, the profiles from adjacent pad rows are nearly aligned in timestamp. In the tilted configuration, the profile in the neighbouring row is shifted to later timestamps. This is consistent with the fact that, for a track inclined in the drift plane, consecutive pad rows correspond to neighbouring portions of the same track with different average drift distances. The comparison therefore illustrates why the internal time structure of the cluster can carry useful geometrical information for inclined-track reconstruction.

To address this, a sub-centroid refit method was developed. The idea is to
replace the single centroid in each pad row by several sub-centroids, so that the time information within a pad row is used more directly in the track fit.
The procedure is as follows:
\begin{enumerate}
    \item Loop over the pad rows of a reconstructed track.
    \item For the selected pad row, remove its centroid and refit the remaining track.
    \item Use this leave-one-out refit to estimate the local track slope in the \(yz\) plane, \(dz/dy\).
    \item For each ADC sample in the selected pad row, estimate its sub-row position from
    \begin{equation}
    y_i = y_{\mathrm{center}} + \frac{z_i - z_{\mathrm{ref}}}{dz/dy},
    \label{eq:subcentroid_yi}
    \end{equation}
    where \(y_{\mathrm{center}}\) is the nominal centre of the pad row and \(z_{\mathrm{ref}}\) is a reference drift coordinate for the row.
    \item Split the ADC samples into \(N_{\mathrm{sub}}\) intervals within the pad row.
    \item Construct ADC-weighted sub-centroids in these intervals and use them in the updated refit.
\end{enumerate}

Several definitions of the sub-centroid intervals were tested. In the
equal-\(y\) split, the ADC samples are first mapped to local positions within the pad row using Eq.~\ref{eq:subcentroid_yi}. The pad-row segment is then divided into intervals of equal length in \(y\), and one ADC-weighted sub-centroid is constructed in each interval. In the equal-\(z\) split, the samples are divided into intervals of equal drift-coordinate width. In the equal-charge split, the interval boundaries are chosen so that each sub-centroid contains approximately the same integrated ADC charge. A final case was also tested in which each ADC timestamp defines one sub-centroid.

\begin{figure}[H]
  \centering
  \begin{subfigure}[t]{0.49\textwidth}
    \centering
    \includegraphics[width=\linewidth]{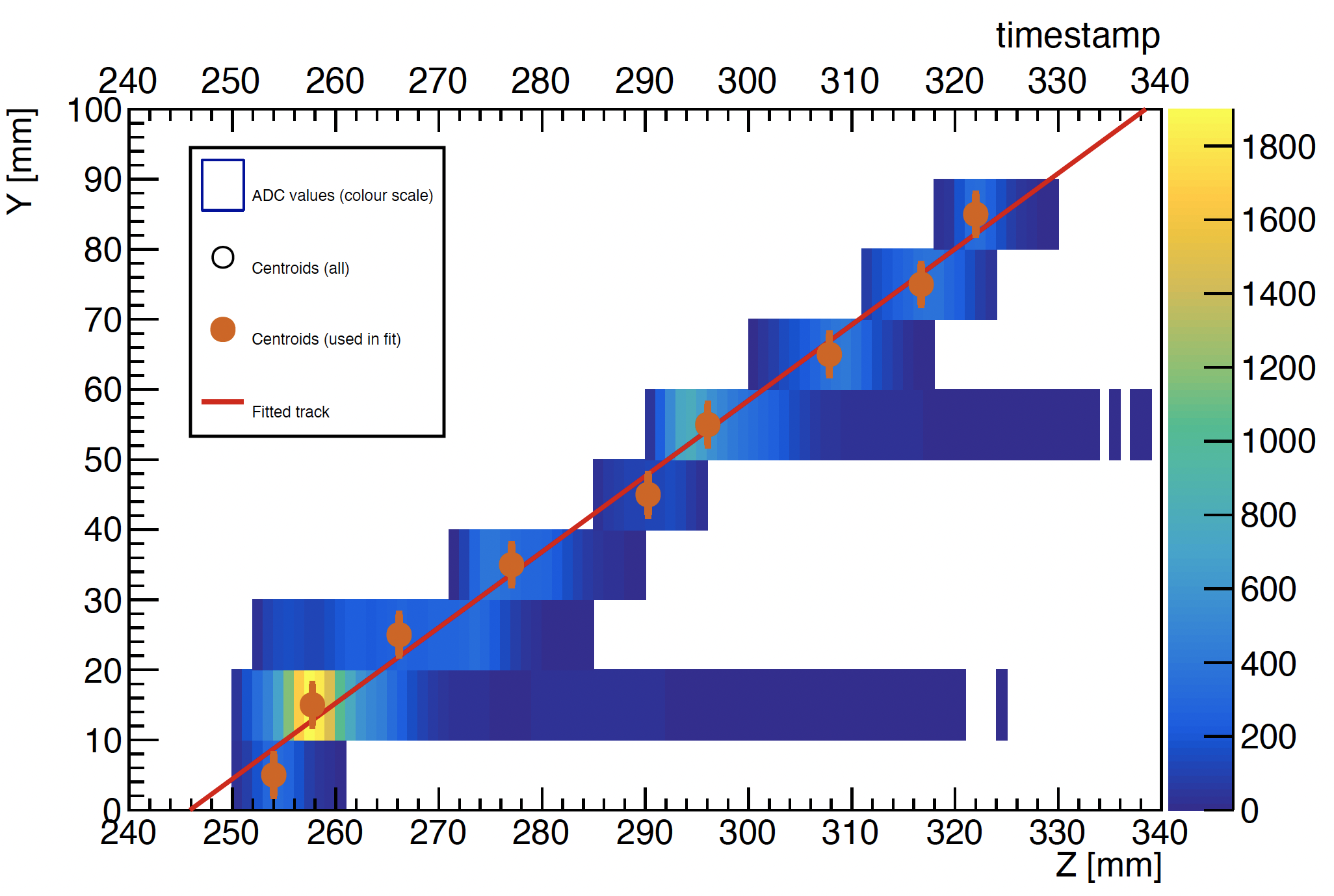}
    \caption{Before the sub-centroid refit.}
    \label{fig:example_track_run007_before_refit}
  \end{subfigure}
  \hfill
  \begin{subfigure}[t]{0.49\textwidth}
    \centering
    \includegraphics[width=\linewidth]{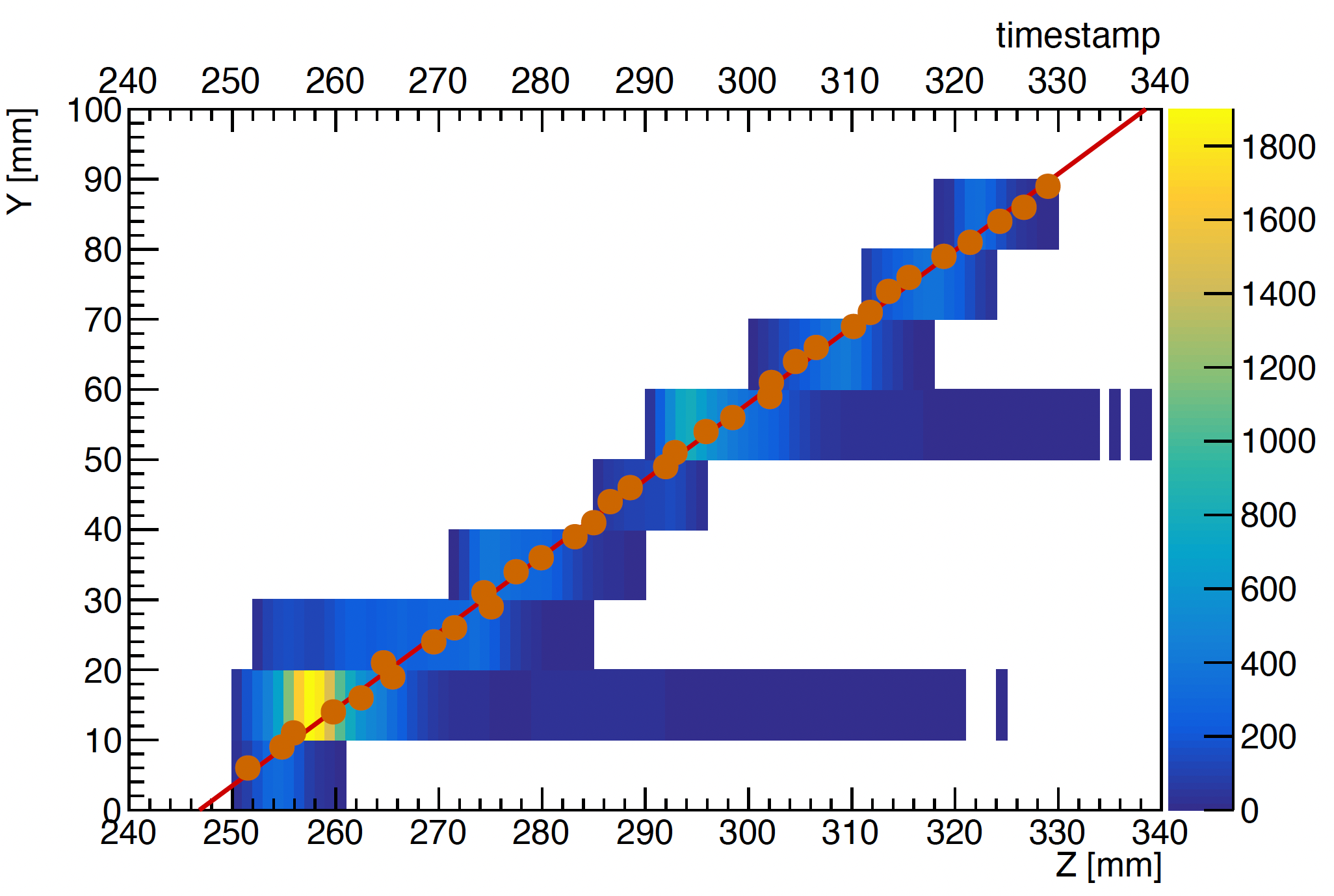}
    \caption{After the sub-centroid refit.}
    \label{fig:example_track_run007_after_refit}
  \end{subfigure}
  \caption{Example of an inclined CCB track before and after the sub-centroid refit. Before the refit, each pad row contributes one centroid at the row centre in \(y\). After the refit, four equal-\(y\) sub-centroids per pad row are reconstructed.}
  \label{fig:run7_example_track_before_after}
\end{figure}

An example of the effect of this procedure on a reconstructed inclined track is shown in Figure~\ref{fig:run7_example_track_before_after}. Before the
refit, each pad row contributes one centroid located at the pad-row centre in
\(y\). After the refit, the same track is represented by sub-centroids. The
fitted track describes the internal time structure of the cluster more closely,
indicating that the stretched time profile is represented more accurately.

Two related residual quantities were used to evaluate the method. The ordinary leave-one-out residual width, used elsewhere in this analysis, is denoted
\(\sigma^{\mathrm{LOO}}_z\). It is obtained from residuals in the \(z(y)\)
projection, where one fitted point is removed, the track is refitted, and the removed point is predicted:
\begin{equation}
r^{\mathrm{LOO}}_{z,i}
=
z_i - z_{\mathrm{fit},i}(y_i).
\end{equation}
The Gaussian width of the resulting residual distribution is denoted
\(\sigma^{\mathrm{LOO}}_z\).

When a pad row is split into several sub-centroids, the resulting points are not fully independent because they originate from the same original cluster. Therefore, an additional row-level validation quantity was used. For this test, all sub-centroids originating from one original pad row are removed
together, the track is refitted, and the removed row is predicted from the
remaining rows. The residual is defined as
\begin{equation}
r^{\mathrm{row\,LOO}}_z =
z_{\mathrm{row}} - z_{\mathrm{fit},\mathrm{row}}(y_{\mathrm{row}}),
\end{equation}
and the Gaussian width of the resulting residual distribution is denoted
\begin{equation}
\sigma^{\mathrm{row\,LOO}}_z .
\end{equation}
This quantity tests how well the track predicts a pad row that was not used in the fit. It is therefore used as the main row-level validation quantity for the sub-centroid study.

Table~\ref{tab:subcentroid_refit_summary} summarises the tested pad row
splitting definitions. The equal-\(y\) split was scanned from two to six
sub-centroids per pad row. The equal-\(z\), equal-charge and one-sub-centroid-per-timestamp constructions are included to test whether a different way of subdividing the pad rows gives better row-level prediction or track-level fit quality.

\begin{table}[H]
  \centering
  \scriptsize
  \renewcommand{\arraystretch}{1.15}
  \setlength{\tabcolsep}{2.6pt}
  \caption{Comparison of sub-centroid configurations for inclined CCB tracks. The equal-\(y\) split was scanned for different numbers of sub-centroids. The equal-\(z\) and equal-charge rows use four sub-centroids for direct comparison with the retained equal-\(y\) configuration. The timestamp split constructs one sub-centroid for each ADC timestamp of the pad-row pulse.}
  \label{tab:subcentroid_refit_summary}

  \makebox[\textwidth][c]{%
  \begin{tabular}{l c c c c c c c}
    \hline
    Method &
    Subdivision &
    \shortstack{\(\sigma^{\mathrm{row\,LOO}}_x\)\\{[}mm{]}} &
    \shortstack{\(\sigma^{\mathrm{LOO}}_x\)\\{[}mm{]}} &
    \shortstack{\(\sigma^{\mathrm{row\,LOO}}_z\)\\{[}mm{]}} &
    \shortstack{\(\sigma^{\mathrm{LOO}}_z\)\\{[}mm{]}} &
    \shortstack{median RMS\\\(yz\) [mm]} &
    \shortstack{median\\\(\chi^2_\nu(yz)\)} \\
    \hline
    Standard centroids & -- & -- & 0.590 & -- & 1.301 & 0.996 & 0.185 \\
    Equal-\(y\) split & 2 sub-centroids & 0.570 & 0.527 & 0.482 & 0.633 & 0.552 & 0.351 \\
    Equal-\(y\) split & 3 sub-centroids & 0.574 & 0.516 & 0.440 & 0.826 & 0.727 & 0.579 \\
    Equal-\(y\) split & 4 sub-centroids & 0.576 & 0.508 & 0.435 & 0.549 & 0.534 & 0.305 \\
    Equal-\(y\) split & 5 sub-centroids & 0.575 & 0.499 & 0.441 & 0.590 & 0.620 & 0.406 \\
    Equal-\(y\) split & 6 sub-centroids & 0.575 & 0.492 & 0.429 & 0.624 & 0.736 & 0.566 \\
    Equal-\(z\) split & 4 sub-centroids & 0.609 & 0.586 & 0.541 & 0.623 & 1.376 & 2.023 \\
    Equal-charge split & 4 sub-centroids & 0.563 & 0.620 & 0.482 & 0.565 & 0.571 & 0.349 \\
    Timestamp split & one sub-centroid per timestamp & 0.605 & -- & 0.515 & 0.614 & 1.644 & 2.725 \\
    \hline
  \end{tabular}%
  }
\end{table}

The results show that splitting the pad-row signal into sub-centroids with the equal $y$ split improves the reconstruction of inclined tracks in the \(yz\) projection. For the standard centroid reconstruction, \(\sigma^{\mathrm{LOO}}_z = 1.301~\mathrm{mm}\). With four equal-\(y\) sub-centroids per pad row,
\(\sigma^{\mathrm{LOO}}_z\) decreases to \(0.549~\mathrm{mm}\), and the
row-level validation gives \(\sigma^{\mathrm{row\,LOO}}_z =
0.435~\mathrm{mm}\). The improvement shows that the internal time structure of the pad-row signal contains useful position information for inclined tracks.

The comparison of the splitting definitions shows that the geometrical
division in \(y\) gives the best overall compromise. For the same number of sub-centroids, \(N_{\mathrm{sub}}=4\), the equal-\(z\) split gives a much larger median RMS residual and median \(\chi^2_\nu\), while the equal-charge split gives a larger \(\sigma^{\mathrm{row\,LOO}}_z\) than the equal-\(y\) split and does not improve the track-level quantities.

Within the equal-\(y\) scan, six sub-centroids give the smallest
\(\sigma^{\mathrm{row\,LOO}}_z\), but the difference relative to four
sub-centroids is only \(0.006~\mathrm{mm}\). The four-sub-centroid case gives the best track-level behaviour among the scanned equal-\(y\) configurations,
with the smallest \(\sigma^{\mathrm{LOO}}_z\), median RMS residual and median
\(\chi^2_\nu\). Four equal-\(y\) sub-centroids per pad row were therefore
chosen as the preferred configuration for the inclined-track refit, based on
the balance between row-level validation and track-level fit quality.

The one-sub-centroid-per-timestamp construction gives one sub-centroid for each ADC timestamp of the pulse in a pad row. This tests the limiting case in which the time structure is used with the finest available subdivision. Such a fine split reduces the number of ADC samples contributing to each point, so larger residual widths are not unexpected. However, the results do not show a compensating improvement in the fitted-track quality: the row-level width remains larger than for the equal-\(y\) configurations with three to six sub-centroids, while the median RMS residual and median \(\chi^2_\nu\) are substantially worse. This indicates that the time structure of the pad-row
signal is useful, but that treating individual ADC timestamps as independent
spatial measurements over-divides the pulse.

\FloatBarrier

\section{Discussion and Outlook for the HIBEAM TPC}
\label{sec:discussion_outlook}

Applying the reconstruction algorithm to the experimental data showed that the prototype TPC can reach residual widths around \(0.5~\mathrm{mm}\), provide
stable charge sharing on the zigzag pad plane, and give a nearly uniform
\(dE/dx\) response through the drift volume. Overall, the tracking performance was mostly stable across the
cosmic measurements and the proton elastic-scattering measurements, although
specific challenges were observed for tracks inclined in the \(yz\) plane and
for events with overlapping clusters or higher track multiplicity.

For the \(yx\) projection, the experimental residual widths remain close to
\(0.5~\mathrm{mm}\) among the different experimental data sets. This suggests
that the reconstruction in this projection is now largely limited by the width
of the readout pads. The zigzag pattern improves the reconstruction of the
\(x\) coordinate compared with rectangular pads of the same
rectangular-equivalent size, because it provides charge sharing between
neighbouring pad columns. However, a substantial further improvement in the
\(yx\) residual width would require a reduced zigzag-pattern width.

For the \(yz\) projection, the residual widths are also affected by the detector
geometry, but there is still room for improvement in how the information in the
cluster time profile is used. For tracks inclined in the \(yz\) plane, the signal
associated with one pad row extends over a longer drift time. The extracted ADC
time profile therefore covers a longer interval within the pad-row signal and
becomes more sensitive to local ionisation fluctuations. These fluctuations
deteriorate the time estimate for the reconstructed
drift coordinate of centroid. The sub-centroid refit technique was introduced to reduce this effect by
using more of the information contained in the extended time profile of cluster, instead of
describing the whole cluster by one centroid placed in the middle of pad row.

In this sub-centroid refit technique, the ADC values from one cluster are split into several sub-centroids along the track inside the pad row. This takes
advantage of the time information in the extended cluster profile, rather than averaging the whole pad-row signal into one centroid placed at the centre of
the pad row. With four equal-\(y\) sub-centroids per pad row, the
\(yz\)-projection residual width was reduced substantially, showing that this
novel technique can improve the reconstruction of the drift coordinate for
inclined tracks.

The studies of operation parameters of TPC showed that the reconstruction performance
depends on a balance between gain, charge sharing, ADC saturation and drift
velocity. In the case of the cosmic data, a GEM
voltage of \(V_{\mathrm{GEM}} = 350~\mathrm{V}\) per GEM provided a good
compromise with a clean \(dE/dx\) distribution, stable residual widths and
sufficient charge sharing between neighbouring pads, while avoiding the saturation observed at higher gain. The gas mixture and operating
point were chosen to give freedom for later optimisation in the direction
indicated by the prototype tests and by the finalisation of the performance
requirements from full HIBEAM simulations.

The cosmic measurements are especially relevant because cosmic-muon tracks provide a minimum-ionising reference with track topology and
ionisation levels similar to the charged pions that the final HIBEAM TPC aims
to reconstruct. For this dataset, a global 70th-percentile threshold was
determined from the centroid \(dE/dx\) distribution and used to calculate a
truncated-mean \(dE/dx\) value for each reconstructed track. This reduces the
influence of large local \(dE/dx\) fluctuations, for example from rare large
energy transfers or \(\delta\)-electrons, which would otherwise broaden the
track-level arithmetic mean. The resulting truncated-mean distribution is
narrower and closer to a Gaussian shape, with a relative width of about
\(16\%\). This should be understood as an indicative result for identifying
minimum-ionising tracks in the TPC.

The proton elastic-scattering measurements provided an opportunity to
study a large number of tracks entering the TPC around a given angle, while
also testing the TPC under conditions with higher ionisation than in
the cosmic measurements. For the perpendicular configuration, the selected
trigger tracks were reconstructed with residual widths similar to those obtained
in the cosmic measurements. The challenge was to identify the trigger particle
coming from the proton elastic-scattering event: the triggered protons and
deuterons had to pass through about \(1~\mathrm{cm}\) of plastic scintillator,
corresponding to a threshold of about \(31~\mathrm{MeV}\) per nucleon, while
there could be other background tracks arriving within the trigger time window too.
These lower-energy background tracks produce larger ionisation in the TPC
and can interfere with the triggered tracks. For the higher-multiplicity perpendicular data, the larger number of
reconstructed tracks increased the chance that clusters from background tracks
interfered with the trigger-track clusters. Such interference can distort the
reconstructed centroid positions and worsen the reconstruction performance.

Lastly, the proton elastic-scattering setup with the TPC tilted by
\(45^\circ\) provided high-statistics data for tracks inclined in the
\(yz\) plane. This made it possible to test the sub-centroid refit technique
described above for improving the reconstruction of the drift coordinate. The
technique proved promising: with four equal-\(y\) sub-centroids per pad row,
the \(yz\)-projection residual width was reduced from
\(1.30~\mathrm{mm}\) to \(0.55~\mathrm{mm}\).

The simulated crossing-track data provided a complementary stress test of the
tracking algorithm. A large fraction of events with two crossing tracks could
be reconstructed, while cases with small relative angles between the tracks in
the \(yz\) projection were more difficult, since clusters from different tracks
then overlapped over several pad rows. Nevertheless, such track topologies are
expected to be rare in the final HIBEAM TPC.

The current TPC prototype offers drift and track lengths close to those
expected for the final HIBEAM TPC, but it does not fully reproduce the field geometry of the final detector. In particular, the prototype has a lateral extension of the drift field larger than the GEM stack in the readout plane, so boundary effects close to the field edge were not fully realistic. Therefore, ongoing work on the next-generation TPC prototype focuses on constructing a field cage closer to the HIBEAM geometry, with a curved shape, in order to study how drift-field distortions close to the detector boundaries affect the track image.

\section{Conclusions}
A TPC prototype, developed for the future HIBEAM annihilation detector, was
characterised using cosmic-muon measurements and proton elastic-scattering data from the experiment conducted at CCB. The work combined characterisation of the TPC prototype with the
development of a track-reconstruction and analysis toolkit. The characterisation focused on the \(dE/dx\) response, spatial resolution from residual widths, estimating the magnitude of charge sharing between pads and the dependence on operating
conditions such as GEM voltage or drift field. The prototype reached sub-millimetre residual widths, showed stable charge sharing from the zigzag pad geometry, and provided an indicative track-level truncated-mean \(dE/dx\) measurement for identifying
minimum-ionising tracks in future \\ particle-identification studies.

The results from the TPC prototype provide a first validation of the basic
design choices and operating conditions within the constraints of the HIBEAM TPC concept. In addition, they provide validation of the tracking algorithm and the analysis toolkit developed in this work. The CCB data extended the validation to tracks from proton-scattering events with substantially higher ionisation than the cosmic-muon measurements, while the tilted configuration showed that the sub-centroid refit can exploit
the internal time structure of clusters to improve the reconstruction of tracks
inclined in the \(yz\) plane. These results also form the basis for a new TPC prototype with more realistic curved field boundaries, which will allow detailed studies of boundary effects and electric-field non-uniformities. Together, the results move the HIBEAM TPC concept closer to a full annihilation detector design for a future neutron--antineutron oscillation search at ESS.

\newpage

\bibliographystyle{elsarticle-num}
\bibliography{references}  

\end{document}